\documentclass[aps,prd,onecolumn,superscriptaddress,nofootinbib,showpacs]{revtex4}

\usepackage{epsfig,amsfonts,amsthm}
\usepackage{amsmath,amssymb}
\usepackage{epstopdf}

\newcommand{\be}{\begin{equation}}
\newcommand{\ee}{\end{equation}}
\newcommand{\ba}{\begin{align}}
\newcommand{\ea}{\end{align}}
\newcommand{\bea}{\begin{eqnarray}}
\newcommand{\eea}{\end{eqnarray}}
\newcommand{\no}{\nonumber\\}

\def\lsim{\mathrel{\rlap{\lower4pt\hbox{\hskip1pt$\sim$}}
    \raise1pt\hbox{$<$}}}         
\def\gsim{\mathrel{\rlap{\lower4pt\hbox{\hskip1pt$\sim$}}
    \raise1pt\hbox{$>$}}}         

\begin{document}

\title{Metastability bounds on the two Higgs doublet model}

\author{A. Barroso}
\affiliation{Centro de F\'{\i}sica Te\'{o}rica e Computacional,
    Faculdade de Ci\^{e}ncias,
    Universidade de Lisboa,
    Av.\ Prof.\ Gama Pinto 2,
    1649-003 Lisboa, Portugal}
\author{P.M.~Ferreira}
    \email[E-mail: ]{ferreira@cii.fc.ul.pt}
\affiliation{Instituto Superior de Engenharia de Lisboa - ISEL,
	1959-007 Lisboa, Portugal}
\affiliation{Centro de F\'{\i}sica Te\'{o}rica e Computacional,
    Faculdade de Ci\^{e}ncias,
    Universidade de Lisboa,
    Av.\ Prof.\ Gama Pinto 2,
    1649-003 Lisboa, Portugal}
\author{I.P.~Ivanov}
    \email[E-mail: ]{igor.ivanov@ulg.ac.be}
\affiliation{IFPA, Universit\'{e} de Li\`{e}ge, All\'{e}e du
6 Ao\^{u}t 17, b\^{a}timent B5a, 4000 Li\`{e}ge, Belgium}
\affiliation{Sobolev Institute of Mathematics, Koptyug avenue 4, 630090, Novosibirsk, Russia}
\author{Rui Santos}
    \email[E-mail: ]{rsantos@cii.fc.ul.pt}
\affiliation{Instituto Superior de Engenharia de Lisboa,
	1959-007 Lisboa, Portugal}
\affiliation{Centro de F\'{\i}sica Te\'{o}rica e Computacional,
    Faculdade de Ci\^{e}ncias,
    Universidade de Lisboa,
    Av.\ Prof.\ Gama Pinto 2,
    1649-003 Lisboa, Portugal}
%

\date{\today}

\begin{abstract}
In the two Higgs doublet model, there is the possibility that the vacuum
where the universe resides in is metastable. We present the tree-level bounds on the
scalar potential parameters which have to be obeyed to prevent that situation.
Analytical expressions for those bounds are shown for the most used potential,
that with a softly broken $Z_2$ symmetry. The impact of those bounds on the model's
phenomenology is discussed in detail, as well as the importance of the current
LHC results in determining whether the vacuum we live in is or is not stable.
We demonstrate how the vacuum stability bounds can be obtained for the most
generic CP-conserving potential, and provide a simple method to implement
them.
\end{abstract}


\maketitle

The two Higgs doublet model (2HDM)~\cite{Lee:1973iz} is one of the simplest extensions of the
Standard Model (SM) of particle physics. It has a rich phenomenology, allowing for the
possibility of spontaneous CP breaking, as a possible explanation for the matter-antimatter
asymmetry in the universe. It has a richer scalar content, with two CP-even scalar particles, a
pseudoscalar and a pair of charged scalars; possible dark mater candidates; and many other
interesting features. For a recent review, see~\cite{Branco:2011iw}. In light of the recent
discovery at the LHC of a particle that closely resembles the SM Higgs boson~\cite{:2012gk,:2012gu},
but which seems to show some deviations from its expected behaviour, we can finally use the
experimental data to choose from the plethora of proposed SM extensions. In particular,
the 2HDM has shown to be quite capable of reproducing the available experimental
results~\cite{Chen:2013kt,Belanger:2012gc,Chang:2012ve,Ferreira:2011aa}.

The price to pay for the rich phenomenology of the 2HDM is a scalar potential which
is much more complex than the SM's, and which possesses a greater number of free parameters.
Although unknown, those parameters are not wholly unconstrained. For instance, to ensure
the existence of a minimum in the theory, the 2HDM scalar potential needs to be bounded from below.
This severely constrains the quartic scalar couplings of the theory~\cite{Deshpande:1977rw}.
It is also usually required that all amplitudes involving scalars preserve perturbative unitarity~\cite{Kanemura:1993hm,Akeroyd:2000wc} - an analogue, for the 2HDM, of the Quigg-Thacker constraints~\cite{Lee:1977yc,Lee:1977eg}.
These bounds, once again, strongly restrict the potential's parameters. Known
experimental evidence also comes into play: electroweak precision data is used, via
the S, T and U observables~\cite{Peskin:1991sw,STHiggs,lepewwg,gfitter1,gfitter2},
to impose bounds on the 2HDM parameter space; and
measurements from B-physics experiments impose serious restrictions on the scalar-fermion
couplings of the 2HDM.

In the present work we will address a new class of bounds on the 2HDM scalar potential, related
to the existence of a metastable neutral vacuum. The vacuum structure of the 2HDM is much richer
than the SM's. In fact, for certain choices of
parameters, it is possible to have a vacuum which spontaneously breaks CP invariance - indeed,
that is the reason Lee first proposed the model in 1973. For other regions of parameter space
the vacuum may break the electromagnetic $U(1)$ symmetry - such vacua are to be avoided at all
costs. And of course, for large regions of parameter values, the vacuum of the 2HDM is ``normal" and breaks
electroweak gauge symmetry but preserves both electromagnetic and CP symmetries. The 2HDM has however a
final surprise
in store: it is possible for the scalar potential to display {\em two} such ``normal" minima, both of them
breaking exactly the same symmetries. Then the possibility arises that one of those minima
is the one where we currently live, where the scalars' vacuum expectation values (vevs) give elementary
particles
their known masses; but in the second, deeper, minimum the vevs are such that all particle masses are
completely different. ``Our" vacuum is then but a metastable one, and the state of lowest energy
lies below us. We call this situation a ``panic" vacuum. In a recent work~\cite{Barroso:2012mj} we
studied under what conditions this occurred, for a specific version of the 2HDM, namely
a softly broken Peccei-Quinn potential~\cite{Peccei:1977hh}. The conclusion reached therein was remarkable:
the current LHC results enable us to conclude that, for this version of the 2HDM, our vacuum is not metastable,
{\em i.e.} it is the potential's global minimum. And this
regardless of any cosmological considerations.

In the present work we will present in detail the conditions under which the most general
CP-conserving 2HDM potential can develop two minima. And also what bounds one can impose to prevent the
occurrence of a panic vacuum. We will analyse in detail the possibility of panic vacua in
the most popular version of the 2HDM, namely the one where a $Z_2$ symmetry has been imposed
on the lagrangian, but that symmetry is softly broken. We will present the exact analytical expressions
for the bounds one needs to impose to ensure absolute stability of the vacuum for this model.
We will show what the current LHC results have to say about the status of ``our" minimum in the $Z_2$
potential.

\section{The vacuum structure of the 2HDM}

The most general renormalizable 2HDM scalar potential is written as
\bea
V &=&
m_{11}^2 |\Phi_1|^2
+ m_{22}^2 |\Phi_2|^2
- \left( m_{12}^2 \Phi_1^\dagger \Phi_2 + h.c. \right)
\no & &
+ \frac{1}{2} \lambda_1 |\Phi_1|^4
+ \frac{1}{2} \lambda_2 |\Phi_2|^4
+ \lambda_3 |\Phi_1|^2 |\Phi_2|^2
+ \lambda_4 |\Phi_1^\dagger\Phi_2|^2
\no & &
+ \left[
\frac{1}{2} \lambda_5 \left( \Phi_1^\dagger\Phi_2 \right)^2
+ \lambda_6 |\Phi_1|^2
\left( \Phi_1^\dagger\Phi_2 \right)
+ \lambda_7 |\Phi_2|^2
\left( \Phi_1^\dagger\Phi_2 \right)
+ h.c. \right],
\label{eq:pot}
\eea
where the coefficients $m_{12}^2$, $\lambda_{5,6,7}$ can be complex.
Whereas in the SM there is only one possible type of vacuum - which preserves both CP and
the electromagnetic $U(1)$ gauge symmetry, but breaks $SU(2)_L\times U(1)_Y$, and which we
call the {\em normal} vacuum - in the 2HDM there are {\em three}. The normal vacuum
corresponds to both doublets acquiring real and neutral vevs,
\be
\langle\Phi_1 \rangle_N = {\displaystyle\frac{1}{\sqrt{2}}}\begin{pmatrix} 0 \\ v_1 \end{pmatrix} \; , \;
\langle\Phi_2 \rangle_N = {\displaystyle\frac{1}{\sqrt{2}}}\begin{pmatrix} 0 \\ v_2 \end{pmatrix},
\label{eq:vevn}
\ee
which can, without loss of generality, be taken to be both positive. In the
{\em charge breaking (CB) vacuum} the dublets have vevs given by
\be
\langle\Phi_1 \rangle_{CB} = {\displaystyle\frac{1}{\sqrt{2}}}\begin{pmatrix} 0 \\
c_1 \end{pmatrix} \; , \;
\langle\Phi_2 \rangle_{CB} = {\displaystyle\frac{1}{\sqrt{2}}}\begin{pmatrix} c_2 \\ c_3
\end{pmatrix} ,
\label{eq:vevcb}
\ee
where all the $c_i$ are real. Finally, in vacua which spontaneously break CP, the fields'
vevs have a relative complex phase,
\be
\langle\Phi_1 \rangle_{CP} = {\displaystyle\frac{1}{\sqrt{2}}}\begin{pmatrix} 0 \\ \bar{v}_1
\end{pmatrix} \; , \;
\langle\Phi_2 \rangle_{CP} = {\displaystyle\frac{1}{\sqrt{2}}} \begin{pmatrix} 0 \\ \bar{v}_2 e^{i\theta}
\end{pmatrix}.
\label{eq:vevcp}
\ee
{\em A priori}, all of these different vacua could coexist in the potential, raising the possibility
of tunneling between different minima. However, in~\cite{Ferreira:2004yd,Barroso:2005sm}, it was shown
that this is impossible. If $V_N$
is the value of the potential at a normal stationary point, and $V_{CB}$ its value at a charge breaking
one, it is possible to show that the difference of the potential depths is given by
\be
V_{CB} - V_N\,=\,\left(\frac{m^2_{H^\pm}}{4 v^2}\right)_N\,
\left[(v_1 c_3 - v_2 c_1)^2 + v_1^2 c_2^2\right]\, ,
\label{eq:diffcb}
\ee
where $v^2 = v_1^2 + v_2^2$ and $m^2_{H^\pm}$ is the square of the charged scalar mass, both of these
quantities computed at the normal stationary point. And if that stationary point is in fact a minimum, then
$m^2_{H^\pm} > 0$ and as such $V_{CB} - V_N \,>\,0$ - the normal minimum is guaranteed to be deeper than
the CB stationary point. In~\cite{Ferreira:2004yd,Barroso:2005sm} it was further proven that the existence of the normal minimum implies that the CB extremum is necessarily a saddle point.

A similar result is valid for the comparison of a normal and CP stationary points: the difference
in depths of the potentials is given by~\cite{Ferreira:2004yd,Barroso:2005sm}
\be
V_{CP} - V_N\,=\,\left(\frac{m^2_A}{4 v^2}\right)_N\,
\left[(\bar{v}_2 v_1 \cos\theta - \bar{v}_1 v_2)^2 + \bar{v}_2^2 v_1^2 \sin^2\theta\right]\, ,
\label{eq:diffcp}
\ee
with $m^2_A$ being the pseudoscalar mass at the normal stationary point. Existence of a normal minimum
thus automatically gives $V_{CP} - V_N \,>\,0$ - again, the normal minimum is the deepest. And again,
in this situation the CP stationary point is a saddle point~\footnote{Of course, it is possible to
choose the values of the parameters of the potential to obtain a CP minimum - but the results we are
discussing here imply that for those parameter values no normal minima can ever be found.}.

These results can best be summarized as a simple theorem: {\em no minima of different natures can
coexist in the 2HDM}. In other words, if the reader chose a set of 2HDM parameters such that a
normal minimum exists, there is no need to worry about the existence of a deeper CB or CP minimum.

The keen reader will notice that the theorem mentions minima of different natures, which
raises the question of knowing how many minima of each type can exist in the 2HDM. For the CB and
CP cases, the answer is simple: for a given set of 2HDM parameters, the minimization conditions
admit only {\em one} CB vacuum of the form of eq.~\eqref{eq:vevcb}, and only {\em one} CP vacuum of
the form of eq.~\eqref{eq:vevcp}~\footnote{Two solutions of the minimization conditions which are related
to one another by gauge transformations are degenerate and as such taken as a single vacuum. Likewise,
any other exact symmetry of the potential gives rise to the same situation.}.

But in certain situations, the minimization conditions allow for several non-equivalent normal
stationary points. And it was shown~\cite{Ivanov:2006yq,Ivanov:2007de} that in fact {\em two} of those solutions
can be minima.
In other words, other than the normal vacuum with vevs given by eq.~\eqref{eq:vevn}, for which
one as $v_1^2 + v_2^2 = (246$ GeV$)^2$, there exists a second normal minimum $N^\prime$,
with different vevs $\{v^\prime_1 , v^\prime_2\}$. For this second minimum, the sum of the squared
vevs takes a different value, smaller or larger than $(246$ GeV$)^2$. And the two minima are not degenerate, in fact
they verify~\cite{Barroso:2005sm,Barroso:2007rr}
\be
V_{N^\prime} - V_{N}\,=\,\frac{1}{4}\,\left[\left(\frac{m^2_{H^\pm}}{v^2}\right)_{N}
- \left(\frac{m^2_{H^\pm}}{v^2}\right)_{N^\prime}\right]\,
(v_1 v^\prime_2 - v_2 v^\prime_1)^2\, ,
\label{eq:diffn}
\ee
where the quantity $\left(m^2_{H^\pm} / v^2\right)$ is evaluated at both minima, $N$ and
$N^\prime$. This raises the possibility that our minimum, with $v =$ 246 GeV, is not the
deepest one. And in fact, for certain regions of the 2HDM potential, $N^\prime$ is found
to be the global minimum of the model - a minimum where the exact same symmetries have been
broken, but where {\em all elementary particles have different masses}. In that situation
our universe could tunnel to this deeper minimum, with obvious catastrophic consequences.
We call this situation the {\em panic vacuum}.

Before we proceed, let us clarify a common misunderstanding about these coexisting minima.
They are sometimes dismissed out of hand as irrelevant for our understanding of the 2HDM,
since {\em obviously} model-makers can simply choose to start in the global minimum of the
theory and develop perturbation theory from that point. Well, that is not correct: we do
not have the freedom to decide in which of both minima
we are currently at. The point is the following: assume we have access to
the most precise experimental data, from some future collider; this data shows,
without margin for doubt, that there are two CP-even scalars, a pseudoscalar and a charged
one - in other words, the 2HDM describes particle physics. Further, let us assume that, from the data, we can
obtain
all necessary information to reconstruct,
precisely, the potential~\cite{kk} (we will discuss this in further detail in
section~\ref{sec:Z2}).
At this point we have the complete potential and can look at the minimization equations.
It is {\em only then}, after the potential's parameters are locked, that we can verify
whether or not there is a second minimum, and if so if it is deeper than ours, and we are
in the panic vacuum. In other words, we cannot decide to choose the potential's parameters
such that ours is the global minimum; it will be the experimental data which will provide
us with that information.

With that clarification out of the way, let us go back to the coexisting normal minima, and
their difference in depths given by eq.~\eqref{eq:diffn}. Should we worry about the
existence of a deeper minimum than ours? When this issue arose in supersymmetry, concerning
dangerous charge and colour breaking vacua~\cite{Frere:1983ag}, the existence of the deeper minimum
was only considered
problematic if the tunneling time from our minimum to it was found to
be inferior to the age of the universe - only in that case should one exclude the parameters
of the theory which originate both minima as dangerous. In this paper we will present, in
section~\ref{sec:life}, an estimate of the tunneling times between minima (a daunting task at best,
even for the 2HDM). And as we will show, in many  cases the current LHC results are, remarkably,
enough to exclude the existence of panic vacua - thus curtailing the need to compute any
tunneling times, anyway.

Another natural question one might ask concerns the compatibility of the metastable vacuum
with the thermal evolution of the Universe.
Is it natural - or possible at all - that the early hot Universe could end up in
a metastable vacuum after cooling down from electroweak temperatures $T_{EW}$?
Indeed, the thermal fluctuations omnipresent at those high temperatures
would preclude formation of a long-lived region which was not in the true vacuum.
This seemingly bars the Universe from getting stuck in a metastable state.
However, in models with a sufficiently complex vacuum landscape - including the 2HDM -
this description is not fully accurate.
Temperature corrections to the free-energy density of the scalar field can be
such that the relative depth of two coexisting minima changes its sign
at a certain critical temperature $T_c$ {\em significantly below} $T_{EW}$.
In the particular case of the 2HDM, this possibility was mentioned and investigated
in~\cite{thermal}.
That means that, when cooling down from $T_{EW}$, the Universe goes through
electroweak symmetry breaking and then stays in the global minimum
until $T$ drops below $T_c$.
After that, the Universe is in a metastable state, but the temperature is already
too low to activate the thermal transition to the true vacuum and the tunneling rate
is also too weak. This mechanism could be the origin of the panic vacuum today. We
can call it the ``vacuum freeze-out''
in analogy with the well-known freeze-out phenomenon for various particle species
in cosmology.
The only difference is that the origin of the anomalously slow dynamics
which drives the system out of thermal equilibrium is not the Universe expansion,
but rather the tiny dramatic tunneling rate.
Checking which of the panic vacuum points we find below are compatible with
the vacuum freeze-out is a separate issue, which is left for future work.
For the purpose of the present paper, it is sufficient to stress
that this mechanism is present in 2HDM and does not require any fine-tuning.

Eq.~\eqref{eq:diffn} has one major drawback - it is written in terms of the vevs of
both minima. That makes it quite cumbersome to deal with, if one wishes to know whether
one's minimum is the global one of the potential. In fact, both $\{v_1 , v_2\}$
and $\{v^\prime_1 , v^\prime_2\}$ are, in general, the solutions of two coupled
cubic equations.
The ideal situation
would be to have a set of conditions that the potential should obey to prevent
the occurrence of panic vacua, {\em written in terms of quantities pertaining
to the minimum N alone}. This is in fact possible, based on the work of
ref.~\cite{Ivanov:2006yq,Ivanov:2007de,ivanovPRE}.
There, the generic conditions that specificy the existence of two normal minima in the
2HDM were obtained, as well as the means to answer the question of whether a given
minimum is the global one. Those methods were recently applied to a simple version of
the 2HDM, the softly broken Peccei-Quinn potential. In ref.~\cite{Barroso:2012mj} we provided
very simple conditions, trivial to implement, that ensure the non-existence of panic vacua.
We will now  show how those conditions can be obtained for more general models,
and what conclusions about the potential's minimum one can extract from the current
LHC data.

\section{The case of the softly broken $Z_2$ potential}
\label{sec:Z2}

The most used version of the 2HDM, in a vast array of theoretical and phenomenological
applications, has a $Z_2$ symmetry, $\Phi_1 \rightarrow \Phi_1$ and
$\Phi_2 \rightarrow -\Phi_2$. This symmetry was initially introduced to prevent the
occurrence of tree-level flavour-changing Higgs-mediated
interactions~\cite{Glashow:1976nt,Paschos:1976ay}, and
it eliminates, in the potential of eq.~\eqref{eq:pot}, the parameters $m_{12}^2$,
$\lambda_6$ and $\lambda_7$. However, in order to allow for a vaster parameter space
after unitarity constraints are put in, the $Z_2$ symmetry is
softly broken by the re-introduction of the quadratic term proportional to $m_{12}^2$.
In what follows we will consider the case where we have further required that the only source
of CP violation in the model is that of the SM, {\em i.e.} an explicit violation of CP by the
Yukawa terms, originating a complex CKM matrix. As such all coefficients of the potential
are taken as real~\footnote{There is however much interest in the complex 2HDM, in which
we relax this assumption and consider complex values for $m_{12}^2$ and
$\lambda_5$~\cite{Ginzburg:2002wt,Arhrib:2010ju}. See also~\cite{Barroso:2012wz}.}.
Requiring that the potential has a stationary point with vevs
given by eq.~\eqref{eq:vevn} is tantamount to solving the following minimization equations,
\bea
m^2_{11} v_1 - m^2_{12} v_2 &+& \left(\lambda_1 v_1^2 + \lambda_{345} v_2^2\right)\frac{v_1}{2} = 0 \no
m^2_{22} v_2 - m^2_{12} v_1 &+& \left(\lambda_2 v_2^2 + \lambda_{345} v_1^2\right)\frac{v_2}{2} = 0,
\label{eq:min}
\eea
where we have defined $\lambda_{345} = \lambda_3 + \lambda_4 + \lambda_5$. In terms of the
soft-breaking term $m^2_{12}$, the masses of the CP-even scalars - the lightest $m_h$ and
heaviest $m_H$ -, the pseudoscalar mass $m_A$, the charged Higgs mass $m_{H^\pm}$, the
mixing angle $\alpha$ of the CP-even mass matrix and the angle $\beta$ defined as
$\tan\beta = v_2/v_1$, we have the following expressions for the quartic couplings:
\bea
\lambda_1
&=&
\frac{1}{v^2 c_\beta^2}\left(c_\alpha^2 m_H^2 + s_\alpha^2 m_h^2
- m^2_{12}\frac{s_\beta}{c_\beta}\right) , \no
\lambda_2
&=&
\frac{1}{v^2 s_\beta^2}\left( s_\alpha^2 m_H^2 + c_\alpha^2 m_h^2
- m^2_{12}\frac{c_\beta}{s_\beta}\right), \no
\lambda_3
&=&
\frac{2 m_{H^\pm}^2}{v^2} +
\frac{s_{2 \alpha} (m_H^2 - m_h^2)}{v^2 s_{2 \beta}} - \frac{m^2_{12}}{v^2 s_\beta c_\beta},
\no
\lambda_4
&=&
\frac{m_A^2 - 2 m_{H^\pm}^2}{v^2} + \frac{m^2_{12}}{v^2 s_\beta c_\beta}, \no
\lambda_5
&=& \frac{m^2_{12}}{v^2 s_\beta c_\beta} - \frac{m_A^2}{v^2}.
\label{eq:coup}
\eea
Now, all of the quantities which appear in the equations above are, in principle,
possible to measure in experiments. The physical masses can be obtained by looking at
invariant mass peaks. To establish which of the neutral scalars is $A$ we need only
look at their decays to $ZZ$ or $WW$ - $A$ will not have those. The angles $\alpha$
and $\beta$ can be obtained from combined measurements of the decays
of $h$ and other scalars to $ZZ$, $b\bar{b}$ and $\tau^+\tau^-$. Finally, the soft breaking
term $m^2_{12}$ can be extracted, for instance, from a precision measurement of
$h,H\rightarrow \gamma\gamma$.

Of course, even if all of these scalars were discovered,
the LHC almost certainly would not be able to provide enough precision for accurate
determinations of all the $\lambda_i$, but the point we wish to stress is this:
collider experiments are {\em a priori} sufficient to determine all parameters
in eqs.~\eqref{eq:coup} and, from those, the quartic couplings $\lambda_i$. Using
the minimisation conditions~\eqref{eq:min} we would then determine the values of
$m^2_{11}$ and $m^2_{22}$, and {\em the parameters of the potential would be uniquely
determined from experiments.}

At this point, in possession of all of the parameters of the potential, we can go back
to the minimisation conditions~\eqref{eq:min} and try to solve them for different
values of the vevs $v_1$ and $v_2$. The soft breaking term $m^2_{12}$ renders an
analytical solution of these equations impossible. Of course, they allow for the trivial
solution, both vevs equal to zero - a maximum of the potential. But it has been
shown~\cite{Barroso:2005sm,Ivanov:2007de} that they can lead to several solutions, and at most {\em two}
non-degenerate minima. In fact, expressing the charged Higgs mass in terms of the
parameters of the potential (see, for instance, eq. (204) of~\cite{Branco:2011iw}),
we can rewrite eq.~\eqref{eq:diffn} as
\be
V_{N^\prime} - V_{N}\,=\,\frac{m^2_{12}}{4 v_1 v_2}\,
\left(1 - \frac{v_1 v_2 }{v^\prime_1 v^\prime_2}\right)\,
(v_1 v^\prime_2 - v_2 v^\prime_1)^2\, ,
\ee
where once again we see the crucial importance that the soft breaking term has -
without it,  the minima $N$ and $N^\prime$ would be degenerate.

And thus, there is the following tantalizing possibility: in the future,
a precise determination of the parameters of the 2HDM potential leads us, by
solving eqs.~\eqref{eq:min}, to determine that the model has more than one minimum;
and, for some choices of parameters, that the minimum the universe currently resides in
is {\em not} the global one - the panic vacuum we alluded to in the introduction.
We are now going to provide the reader with a set of simple criteria to answer the following
questions: under what conditions can there be two normal minima in the potential? Under what
conditions is our vacuum, with $v = 246$ GeV, not the global one? We will now write those
conditions, postponing their demonstration until section~\ref{sec:dem}.

\subsection{Existence of two minima}

The softly broken $Z_2$ 2HDM potential can have two normal minima if
the two following conditions are met:
\bea
m_{11}^2 + k^2\, m_{22}^2
&<& 0,
\label{eq:M0} \vspace{0.5cm}
\\
\sqrt[3]{x^2} + \sqrt[3]{y^2}
&\leq&  1,
\label{eq:astr}
\eea
where we have defined
\bea
x
&=&
\frac{4\ k\ m_{12}^2}{
m_{11}^2 + k^2\, m_{22}^2}\,
\frac{\sqrt{\lambda_1 \lambda_2}}{
\lambda_{345} - \sqrt{\lambda_1 \lambda_2}}, \vspace{0.5cm}
\nonumber\\
y
&=&
\frac{m_{11}^2 - k^2\, m_{22}^2}{
m_{11}^2 + k^2\, m_{22}^2}\,
\frac{\sqrt{\lambda_1 \lambda_2} + \lambda_{345}}{
\sqrt{\lambda_1 \lambda_2} - \lambda_{345}},
\label{eq:xy}
\eea
and also
\be
k = \sqrt[4]{\frac{\lambda_1}{\lambda_2}}.
\ee
These conditions are necessary and sufficient conditions for the existence of
four~\footnote{In fact they are eight, but since the potential is invariant under $\Phi_1
\rightarrow -\Phi_1$, $\Phi_2 \rightarrow -\Phi_2$, four of them are degenerate with the other four.
This is a manifestation of the $U(1)_Y$ transformation, and it should not be taken into account.}
stationary points in the potential - but do not guarantee that two of those are minima
(see appendix~\ref{ap:min}).
Nonetheless, only under these circunstances can the potential have a maximum of two normal
minima~\cite{Ivanov:2007de}. Notice that
these are trivial extensions of the conditions considered in~\cite{Barroso:2012mj} for the
softly broken $U(1)$ model, which is a particular case of the $Z_2$ case we
are considering here (with $\lambda_5 = 0$). Notice that they can be written only
in terms of the potential's parameters, without any mention of a specific vacuum.

In order to study the importance of the bounds of eqs.~\eqref{eq:M0},~\eqref{eq:astr}, we
have performed a vast scan over the parameter space of the 2HDM. We have
taken $m_h = 125$ GeV, $125 < m_H < 900$ GeV, $90 < m_A, m_{H^\pm} < 900$ GeV,
$-\pi/2 < \alpha < \pi/2$, $1 < \tan\beta < 40$ and $|m^2_{12}| < 900$ GeV$^2$.
We demanded that the quartic couplings of the potential (calculated from
eqs.~\eqref{eq:coup}) obey
\bea
\lambda_1 > 0 & , &  \lambda_2 > 0 \; ,\nonumber \\
\lambda_3 > -\sqrt{\lambda_1 \lambda_2} & , &
\lambda_3 + \lambda_4 - |\lambda_5| > -\sqrt{\lambda_1 \lambda_2} \;,
\label{eq:bfb}
\eea
so that the scalar potential is bounded from below (in the ``strong sense" as
defined in ref.~\cite{heidelberg}). We also required that these quartic couplings are
such that they satisfy perturbative unitarity~\cite{Kanemura:1993hm,Akeroyd:2000wc} and the electroweak
precision constraints stemming from the S, T and U parameters~\cite{Peskin:1991sw,STHiggs,lepewwg,gfitter1,gfitter2}.
Our simulation consists of 700000 ``points", each one corresponding to a different combination
of potential parameters.

\begin{figure}[htb]
\centering
\includegraphics[height=8cm,angle=0]{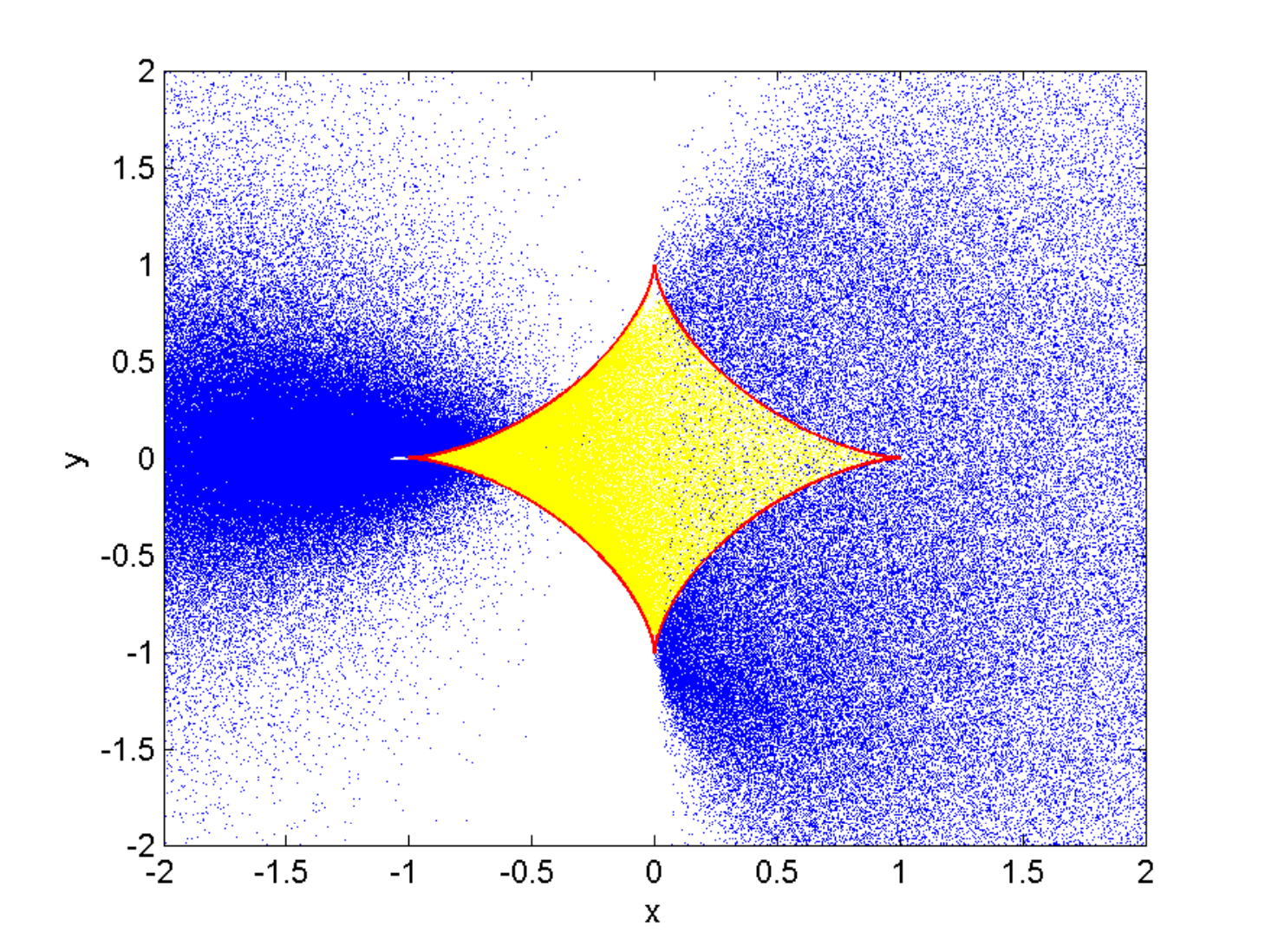}
\caption{Distribution, in the $(x, y)$ plane, of the parameter space
of the $Z_2$ model. Blue (black) represents parameter choices for which
the potential has a single minimum. In yellow (light grey) are the parameters
for which the potential can have two neutral minima. The red (dark grey) curve is the
astroid, inside which points with two minima have to lie.}
\label{fig:ast}
\end{figure}
Eq.~\eqref{eq:astr} defines, in the $(x, y)$ plane, a region of space delimited
by an astroid. All the points with two minima will have to lie inside the astroid and
have $m_{11}^2 + k^2\, m_{22}^2<0$, all other points will necessarily have just one minimum.
In fig.~\ref{fig:ast}
we show how a generic scan of the softly broken $Z_2$ 2HDM includes many points
where two minima are possible, represented with the color yellow (light gray), as opposed to the
points where the minimum is unique, painted blue (black). Though hard to see in
fig.~\ref{fig:ast}, there are some blue points inside the astroid - those for which
the condition~\eqref{eq:M0} is not satisfied. In total, the yellow region consists
of over 140000 points. The existence of two neutral minima is not, therefore, a
curiosity to be dismissed off hand - a full one fifth of the model's parameter space
(after sensible cuts) does not have a single minimum.

\subsection{Existence of a panic vacuum}

We have defined the panic vacuum as the following situation: our vacuum, which
is caracterized by $v = 246$ GeV, is not the global minimum of the potential.
The panic vacua are therefore a subset of the regions of parameter space for which
there are two minima, and the conditions under which that occurs were written in
the previous section.

Remarkably, and again based on the work of
refs.~\cite{Ivanov:2006yq,Ivanov:2007de,ivanovPRE}, it is possible
to write extremely simple conditions for the existence of a panic vacuum. Let us
define the ``discriminant"
\be
D \,=\, m^2_{12} (m^2_{11} - k^2 m^2_{22}) (\tan\beta - k)\,
\label{eq:disc}
\ee
where $\tan\beta = v_2/v_1$ as usual, and written, of course, in terms of the
vevs of ``our" vacuum. The existence of a panic vacuum is thus summarised in the
following theorem:
\be
\mbox{
{\em Our vacuum is the global minimum of the potential if and only if $D > 0$.}
}
\label{eq:cond}
\ee
\begin{figure}[htb]
\centering
\includegraphics[height=8cm,angle=0]{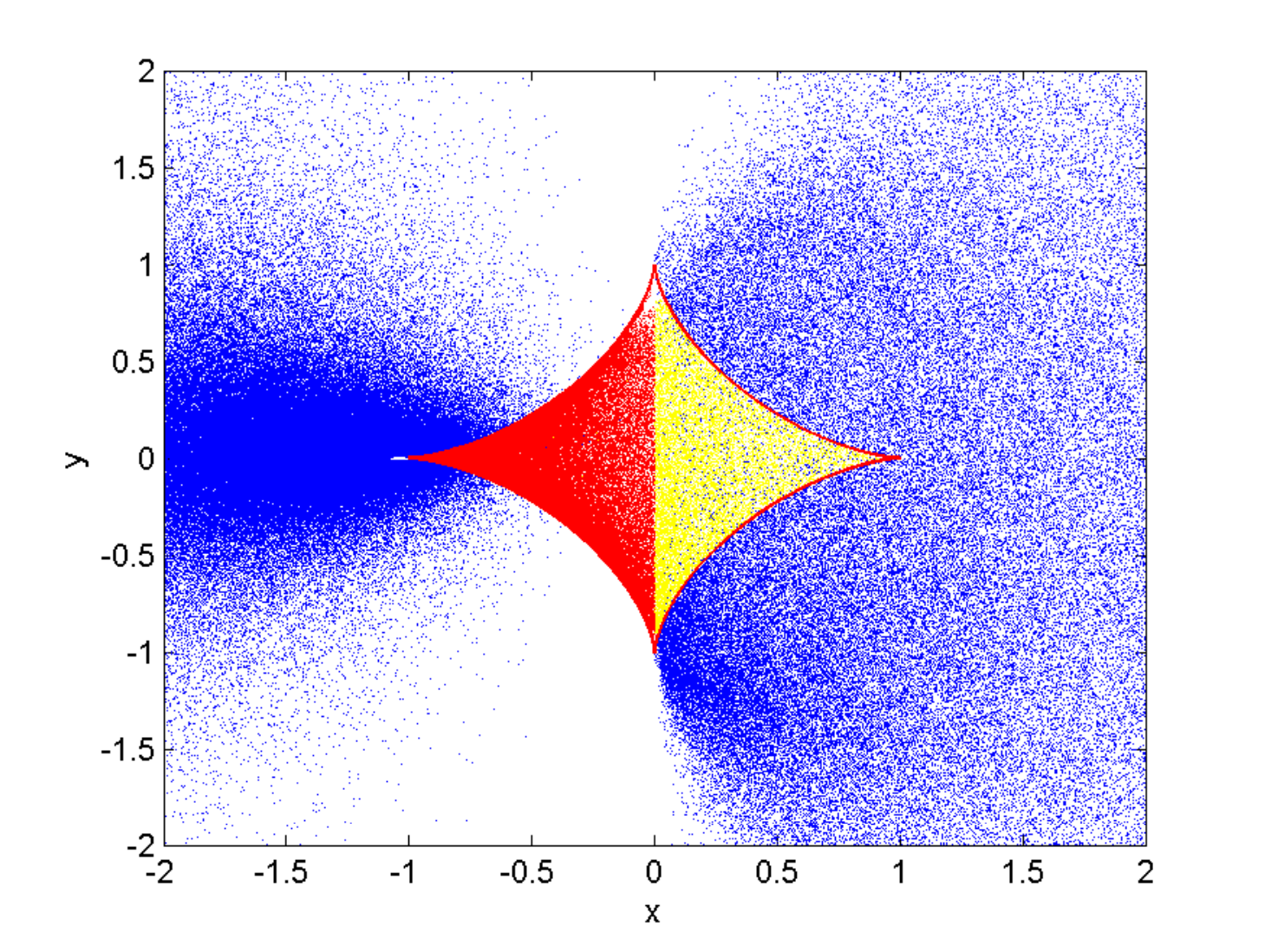}
\caption{Panic vacua in the $(x, y)$ plane. Blue (black) represents points for which
the potential has a single minimum. In yellow (light grey) points corresponding to
two neutral minima. Superimposed on those are red (dark grey) points, which represent panic
vacua. They are bound by the red (dark grey) astroid curve.}
\label{fig:panast}
\end{figure}
Therefore, if we only wish to make certain that we are
in the global minimum of the potential, regardless of the number of those minima,
requiring $D > 0$ is a necessary and sufficient condition. It is not necessary to
verify the conditions shown in the previous section.

Again, the
existence of a minimum of the potential deeper than ours is not just some
curiosity or restricted to a very narrow region of parameter space: in the
scan of 700.000 points of the softly broken $Z_2$ model we discussed in the
previous section, we found almost 126000 points which corresponded to the panic
vacuum scenario! Their distribution in the $(x,y)$ plane, with the variables
defined in eqs.~\eqref{eq:xy}, is shown in fig.~\ref{fig:panast}: the red (dark
grey) points, now superimposed over those shown in fig.~\ref{fig:ast},
correspond to the panic vacua. Notice that only the left-hand side of the astroid
contains panic vacua. This can be demonstrated using the expressions shown in
section~\ref{sec:dem}.

\subsection{LHC results and the existence of panic vacua}

At the time of this writing, the LHC has no hints whatsoever of the existence of more
than one scalar particle. Nonetheless, as we are about to show, the current data can
already tell us a great deal about the nature of our vacuum, and the existence, or lack
thereof, of a deeper minimum in the 2HDM.

First, some explanations concerning what we are comparing with experimental data:
the $Z_2$ symmetry which we impose on the 2HDM scalar potential has to be a symmetry
of the whole lagrangian, otherwise we would be dealing with a non-renormalizable theory.
Thus, that symmetry needs to be extended to the Yukawa sector, and there are a multitude
of ways to do that. In this work we limit ourselves to the two most popular possibilities:
Model I, in which only the doublet $\Phi_2$ couples to fermions; and Model II, in
which $\Phi_2$ couples to all up-type quarks, and $\Phi_1$ couples to all other fermions.
These two models have very different scalar-fermion couplings (see, for instance,
table 2 of~\cite{Branco:2011iw}) and very different phenomenologies.

The  LHC data most relevant for Higgs physics at the moment are the ratios between
observed rates of the Higgs boson decaying into certain particles and their expected
SM values. Assuming that what is being observed is explained by the 2HDM, we
define the said ratio for a given final decay state $f$ of the lightest CP-even
Higgs boson $h$ to be
\be
R_f \,=\, \frac{\sigma^{2HDM} (pp\rightarrow h) \,BR^{2HDM}(h \rightarrow f)}
{\sigma^{SM} (pp\rightarrow h) \,BR^{SM}(h \rightarrow f)},
\ee
including thus both production cross sections $\sigma$ and
the branching ratios (BR) of the Higgs boson. Here, we are considering {\em all}
possible Higgs production mechanisms, but current LHC results already allow us to
distinguish, in certain cases, between some of those. Namely:
\begin{itemize}
\item The gluon-gluon (gg) production mechanism, in which two gluons, one from each
colliding proton, produce a Higgs boson via a triangle of quarks (mostly tops, with a
small percentage of bottoms). Accordingly, we define a $R^{gg}_f$ quantity, considering
only the cross sections of the gg process:
\be
R^{gg}_f \,=\, \frac{\sigma^{2HDM} (pp\rightarrow gg \rightarrow h) \,BR^{2HDM}(h \rightarrow f)}
{\sigma^{SM} (pp\rightarrow gg \rightarrow h) \,BR^{SM}(h \rightarrow f)}.
\ee
\item The vector boson fusion (VBF) mechanism, in which quarks inside the protons radiate
electroweak gauge bosons $V = Z, W$, which ``fuse" to become a Higgs bosons. The VBF
rate is thus defined as
\be
R^{VBF}_f \,=\, \frac{\sigma^{2HDM} (pp\rightarrow VV \rightarrow h) \,BR^{2HDM}(h \rightarrow f)}
{\sigma^{SM} (pp\rightarrow VV \rightarrow h) \,BR^{SM}(h \rightarrow f)}.
\ee
\end{itemize}

Finally, a word on experimental constraints: we already mentioned the electroweak precision
data from LEP that we included in our scan of the 2HDM parameter space, via bounds of the $S$,
$T$ and $U$ observables. Those only provide constraints on the scalar sector of the theory
(assuming no extra generations of fermions are present). But there are plenty of data from
$B$-physics which provide constraints on the fermionic sector of the 2HDM, and which need to be
taken into account. We have used the latest updated bounds from~\cite{nazi}~\footnote{However,
 we have not taken into account the $B_s \rightarrow \mu^+\mu^-$ bound shown in that reference,
 since the analysis presented therein seems specific for the MSSM. Neither did we consider
 the $\bar{B}\rightarrow D^{(*)}\tau^- \bar{\nu}_\tau$ anomaly observed by the BaBar collaboration,
 due to lack of an independent confirmation of it.}. These translate into a limitation
 on the $\tan\beta$-$m_{H^\pm}$ plane.

\begin{figure}[ht]
\centering
\includegraphics[height=6cm,angle=0]{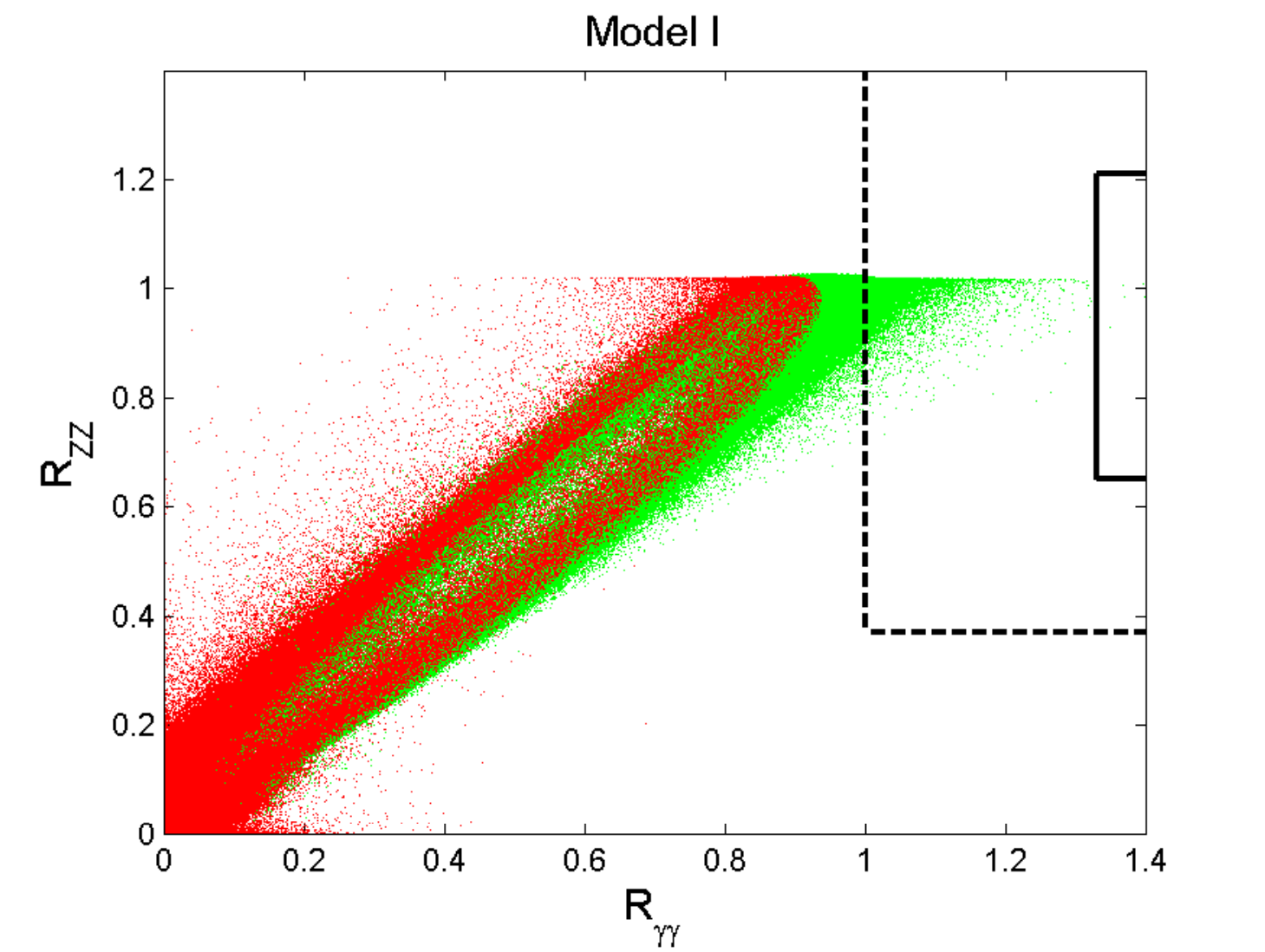}
\includegraphics[height=6cm,angle=0]{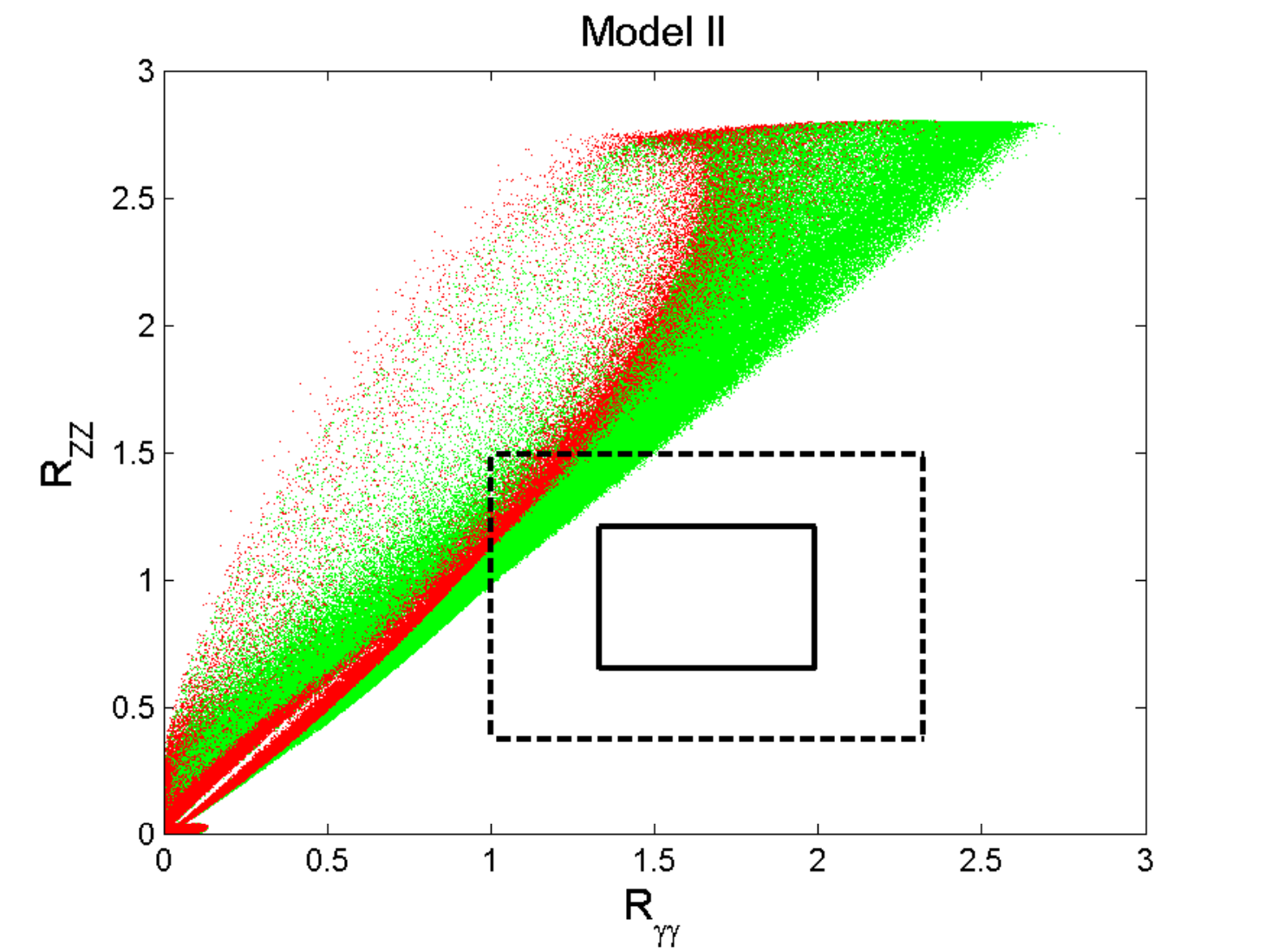}
\caption{$R_{ZZ}$ as a function of $R_{\gamma\gamma}$ for models I and II. Green
(light) points are the total of our simulation, red (dark) ones correspond to
the existence of a panic vacuum. The solid (dashed) lines correspond to
the $1\sigma$ ($2\sigma$) experimental limits.}
\label{fig:zph}
\end{figure}

 In fig.~\ref{fig:zph} we show, for both models considered, the results we obtained
for the rates of the light Higgs $h$ into two $Z$ bosons {\em versus} the rate
of $h$ into two photons. In green (light grey) we show {\em all} the points obtained in
our scan, with the above constraints. In red (black) are the points for which a panic
vacuum occurs. Please notice that the density of points is so great that there are
plenty of green points scattered in the middle of the red ones. Fig.~\ref{fig:zph}
is most relevant for showing that there are regions, in the plane
 $R_{\gamma\gamma}$-$R_{ZZ}$, which are free from panic vacua.
The solid and dashed lines shown in the plots correspond to conservative $1\sigma$
and $2\sigma$ intervals on the combined values for $R_{ZZ}$ and $R_{\gamma\gamma}$,
$R_{ZZ} = 0.93 \pm 0.28$, $R_{\gamma\gamma} = 1.66 \pm 0.33$, which we took from
ref.~\cite{average}.

Notice that we are in the early days of Higgs experimental results, and as such
many of the current numbers (such as the apparent excess in the two-photon rate)
may well change considerably over time. The plots shown in fig.~\ref{fig:zph} are
``invulnerable" to such likely changes, as future, more precise, restrictions on
$R_{\gamma\gamma}$ and $R_{ZZ}$ can be imposed on them quite easily. Still, it
is instructive to consider the current experimental bounds and see that, already,
we can say much about the existence of a panic vacuum.
As we see from fig.~\ref{fig:zph}, for Model I the panic points are well away even from
the $2\sigma$ bands, which include some non-panic region as well. Not so for Model II,
some of the panic region is included at $2\sigma$. Of course,
there are plenty of non-panic vacua choices of parameter space still allowed by the current
data for Model II. As such, in these variables at least, both models seem capable of
describing the current data, but that data does not exclude the possibility, in Model II,
of our vacuum being metastable.

\begin{figure}[ht]
\centering
\includegraphics[height=6cm,angle=0]{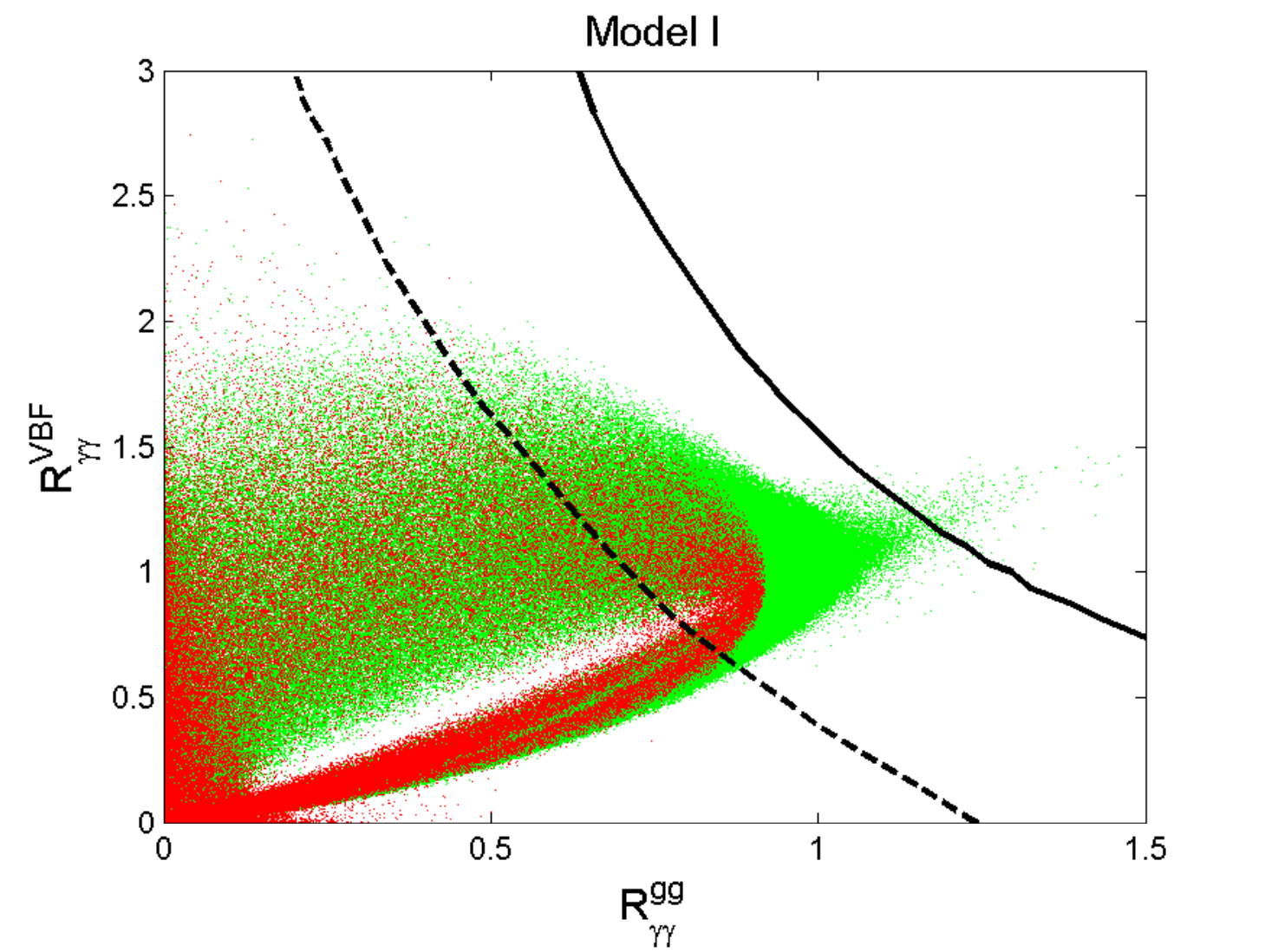}
\includegraphics[height=6cm,angle=0]{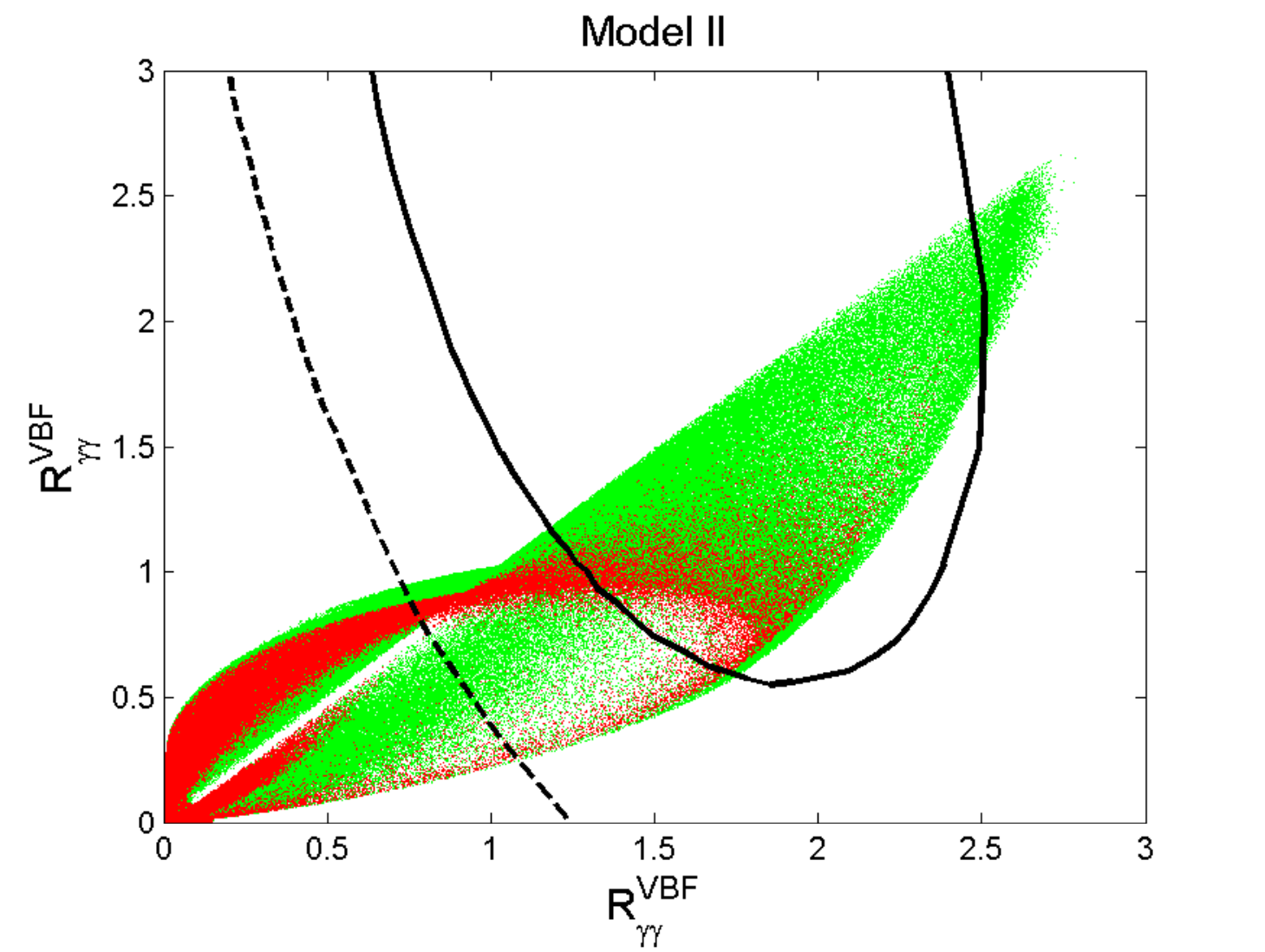}
\caption{$R_{\gamma\gamma}$ including only Higgs production via gluon-gluon fusion,
 and vector boson fusion, for models I and II. Colour code as in fig.~\ref{fig:zph}.}
\label{fig:ggvbf}
\end{figure}

What about other observables, for which there is already considerable information?
The LHC collaborations have been able to measure - with considerable uncertainty -
the ratio of two-photon Higgs events stemming from gluon-gluon production alone,
and from the VBF mechanism alone. They are correlated, and we use the results from the ATLAS
experiment, namely their $1\sigma$ and $2\sigma$ ellipses in the
$R^{gg}_{\gamma\gamma}$-$R^{VBF}_{\gamma\gamma}$ plane~\cite{cms_atlas}. The results of our simulations
are shown in fig.~\ref{fig:ggvbf}, for both Model I and II. In these plots
we see the situation is not as clear-cut as in the previous observables: in Model
I we cannot exclude, at $2\sigma$, the existence of panic vacua~\footnote{Of course,
the panic vacua points which now seem possible have been excluded by fig.~\ref{fig:zph}.};
and in Model II,
even the $1\sigma$ bands include panic vacua solutions. Notice that
the ellipses contain plenty of green/light grey, non-panic points as well. And in these
variables Model II agrees with the data at the $1\sigma$ level, and as such
describes the current data better than Model I.

\begin{figure}[ht]
\centering
\includegraphics[height=6cm,angle=0]{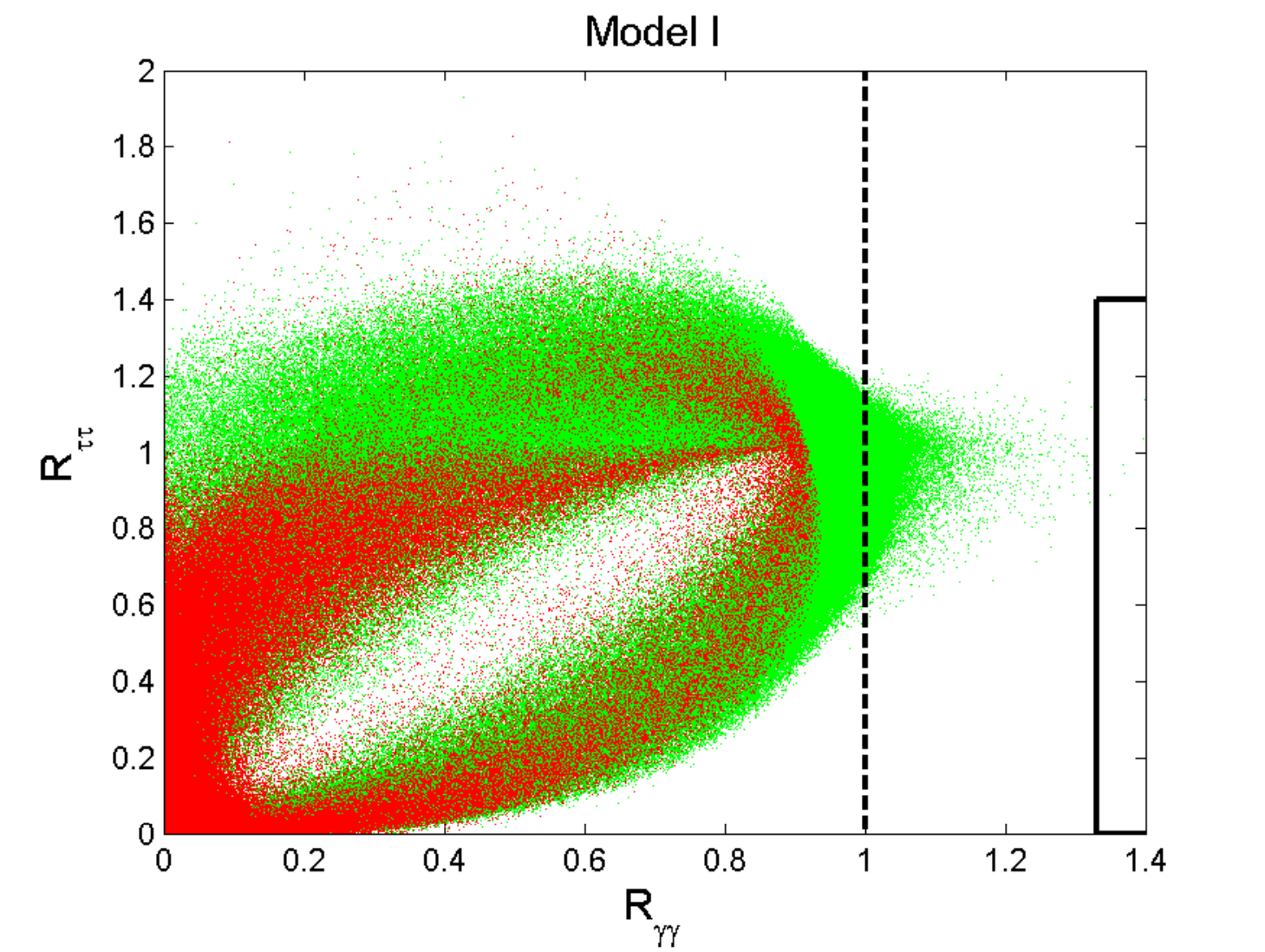}
\includegraphics[height=6cm,angle=0]{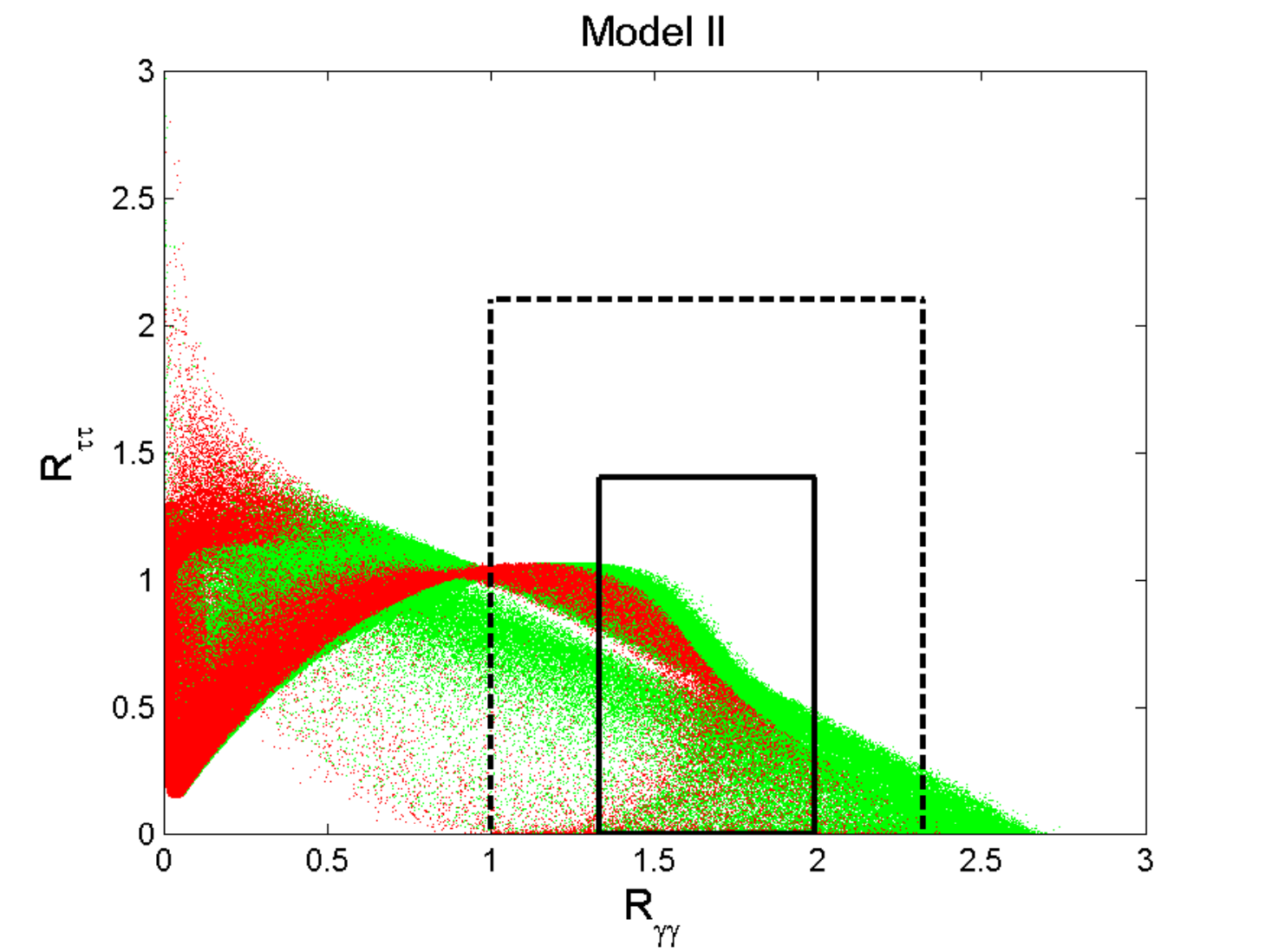}
\caption{$R_{\tau\tau}$ as a function of $R_{\gamma\gamma}$ for models I and II.
Colour code as in fig.~\ref{fig:zph}.}
\label{fig:tauph}
\end{figure}

Finally, to conclude this brief comparison with experimental data, let us look at
the $\tau\tau$ rates, which have recently been measured by both LHC
collaborations~\cite{hcp}. The current results are compatible with the expected
SM value, ATLAS measured $R_{\tau\tau} = 0.7 \pm 0.7$ and CMS,
$R_{\tau\tau} = 0.72 \pm 0.52$. Bearing these numbers in mind, as well as the ones
presented above for the two-photon rates, we present what we have found for
$R_{\tau\tau}$ as a function of $R_{\gamma\gamma}$ in fig.~\ref{fig:tauph}.
The $\tau\tau$ data (we represent the ATLAS bounds, since they are less restrictive),
taken at face value, tells us that panic vacua are at least
$2\sigma$ disfavored in Model I, and the model agrees with the LHC results (at
$2\sigma$, barely, in $R_{\gamma\gamma}$, at $1\sigma$ in $R_{\tau\tau}$).
As for Model II, again at 1-$\sigma$ we notice plenty of panic vacua solutions
not excluded by the data; but for much of Model II's parameter space, we have
agreement with the experimental results at $1\sigma$, with or without panic
vacua.

In short: the current experimental data can already tell us a great deal about
the stability of the vacuum in the 2HDM. For instance, a measurement of $R_{ZZ}$
and $R_{\gamma\gamma}$ very close to 1, with sufficient precision, would exclude the possibility
of panic vacua. Furthermore, this section shows that the same
parameters which produce panic vacua do not correspond to some exotic, uninteresting
corner of the model. They also predict values of observables which are not absurd
and indeed can fall into the bounds of current experimental results. This, we would
argue, is reason enough to take the existence of these panic vacua seriously in
any phenomenological study of the 2HDM.

\section{Panic vacuum bounds for the general CP-conserving potential}
\label{sec:pancp}

The conditions establishing the possible existence of two minima, and in particular of panic
vacua, are amazingly simple and elegant, when written for the softly broken
$Z_2$ model, cf. eqs.~\eqref{eq:M0},~\eqref{eq:astr},~\eqref{eq:disc} and the
statement~\eqref{eq:cond}. They are simpler still for the softly broken Peccei-Quinn
potential, see ref.~\cite{Barroso:2012mj}. But for the most general CP-conserving 2HDM
potential they cannot be written in a concise analytical manner, at least not
in the usual notation. There is, however, a simple ``recipe" which can very
easily be implemented numerically when performing a study of the 2HDM, and in
this section we will give it in detail. We will present the demonstration of all bounds shown
here in section~\ref{sec:dem}. We will leave out of this work any
2HDM potentials with explicit CP-breaking terms, since in those the discussion
is even more difficult. Further, we will write the CP-conserving potential in
a basis where all of its parameters are real. The existence of such a basis is
guaranteed by explicit CP conservation~\cite{guha}.

First, a brief discussion on notation: though writing the 2HDM potential in terms
of doublets, as in eq.~\eqref{eq:pot}, is extremely useful for many calculations
({\em e.g.}, everything dealing with the fermion sector), in some instances a
different notation - in which the potential is written in terms of gauge bilinear
invariants - is crucial. For instance, the comparison of values of potentials at
different vacua (eqs.~\eqref{eq:diffcb}-~\eqref{eq:diffn}) is simple to obtain in
the latter notation, but extremely hard in the former. Likewise, the conditions
for existence of dual minima, or panic vacua, are far easier to establish in the
bilinear formalism, which we now introduce. A remarkable feature of this notation
is the fact that the 2HDM potential has a hidden Minkowski structure, when written
in terms of gauge invariant bilinears. This formalism was developed
in~\cite{Ivanov:2006yq,Ivanov:2007de}. Although similar Minkowskian notation was
used in~\cite{heidelberg,nishi}, those works did not fully exploit the freedom of
non-unitary reparametrization transformations that the Minkowski formalism alludes to.
The gauge invariant bilinears form a covariant 4-vector in a Minkowski
space, $r_\mu$ ($\mu = 0, \ldots\, 3)$, where we define
\be
\begin{array}{rcl}
r_0 &=&
\Phi_1^\dagger \Phi_1 + \Phi_2^\dagger \Phi_2,
\\*[2mm]
r_1 &=&
-\left( \Phi_1^\dagger \Phi_2 + \Phi_2^\dagger \Phi_1 \right)
= -2\,\mbox{Re}\left( \Phi_1^\dagger \Phi_2 \right),
\\*[2mm]
r_2 &=&
 i
\left( \Phi_1^\dagger \Phi_2 - \Phi_2^\dagger \Phi_1 \right)
= -2\,\mbox{Im} \left( \Phi_1^\dagger \Phi_2 \right),
\\*[2mm]
r_3 &=&
-\left( \Phi_1^\dagger \Phi_1 - \Phi_2^\dagger \Phi_2 \right).
\end{array}
\label{eq:rs}
\ee
The allowed vectors $r^\mu  = (r_0, -r_i)$ fill the forward lightcone $LC^+$ defined by $r_0 \ge 0$, $r^\mu r_\mu \ge 0$.
The apex of this cone corresponds to the electroweak symmetric vacuum, its surface corresponds to the
neutral vacua, while its interior corresponds to charge-breaking vacua.
Following the notation
of~\cite{Ivanov:2006yq,Ivanov:2007de} and the conventions of eq.~\eqref{eq:pot}, the
CP-conserving scalar potential
may be written as (with standard Minkowski space conventions)
\be
V =  -M_\mu r^\mu
+ \frac{1}{2} \Lambda_{\mu \nu} r^\mu r^\nu,
\label{eq:potr}
\ee
where the 4-vector $M_\mu$ and the tensor $\Lambda_{\mu \nu}$ are given by
\be
M_\mu = (M_0\,,\,M_i) = \left( -\frac{1}{2}(m_{11}^2 + m_{22}^2),\ \mbox{Re}\left( m_{12}^2 \right),\
0\ ,\ \frac{1}{2}(m_{22}^2 - m_{11}^2) \right)
\label{eq:Mmu}
\ee
(of course, $M^\mu = (M_0\,,\,-M_i)$) and
\be
\Lambda_{\mu \nu} = \frac{1}{2}\left( \begin{array}{cccc}
\displaystyle{\frac{\lambda_1+\lambda_2}{2}} + \lambda_3 &
 \lambda_6 + \lambda_7  & 0 &
\displaystyle{\frac{\lambda_1-\lambda_2}{2}} \\
 \lambda_6 + \lambda_7  &
\lambda_4 + \lambda_5  & 0 &
\lambda_6 - \lambda_7 \\
0 & 0 & \lambda_4 - \lambda_5 & 0 \\
\displaystyle{\frac{\lambda_1-\lambda_2}{2}} &
\lambda_6 - \lambda_7 &
0 & \displaystyle{\frac{\lambda_1+\lambda_2}{2}} - \lambda_3
\end{array} \right).
\label{eq:Lambda}
\ee
As mentioned earlier, we are working in a basis where all parameters are real, which
causes the appearance of several zeros in $\Lambda$ and $M$. With the notation
established, here are the preliminary steps required to verify whether or not the most
general CP-conserving potential can have two neutral minima, and if one of them is
a panic vacuum.
\begin{itemize}
\item The first step in our ``recipe" is the diagonalization of the tensor of the
quartic couplings. Due to the Minkowski indices, this is achieved via a combination
of rotations and Lorentz boosts~\footnote{The diagonalization of $\Lambda_{\mu\nu}$
does not preserve the kinetic terms of the scalars. But it does not affect any details
of calculations of vacua, or the value of the potential at vacua, so we need not worry.}.
But there is a much simpler way, trivial to implement:
define the matrix $\Lambda = {\Lambda_\mu}^\nu$, which is obtained from eq.~\eqref{eq:Lambda}
by simply flipping the sign of the three last columns. The $4\times 4$ matrix $\Lambda$
has eigenvalues $\Lambda_a$ ($a = 0, 1, 2, 3$) determined by the usual equation,
\be
\mbox{det}(\Lambda - \Lambda_a \mathbb{I}) \,=\,0,
\ee
with eigenvectors $V^{(a)}$ which satisfy (no sum in indices)
\be
\Lambda \, V^{(a)} = \,\Lambda_a\, V^{(a)}.
\label{eq:eve}
\ee
Solving for the eigenvectors and eigenvalues of $\Lambda$ is trivially implemented
within any numerical calculation package. Since the matrix $\Lambda$ is not symmetric anymore,
its eigenvalues and eigenvectors are in general complex.
\item The next step is to ensure that the potential is bounded from below. This means
that the eigenvalues of $\Lambda$ must obey the following conditions:
\bea
 & & \mbox{All of them must be real}. \label{eq:bfb1} \\
 & & \Lambda_0 \,> \,0. \\
 & & \Lambda_0 \,> \,\{\Lambda_1\,,\,\Lambda_2\,,\,\Lambda_3\}. \label{eq:bfb3}
\eea
\item The eigenvectors obtained in~\eqref{eq:eve} are then real and can be normalized in such a way that
one of them is time-like, the others space-like. Meaning, if the eigenvector
corresponding to the largest eigenvalue $\Lambda_0$, obtained in~\eqref{eq:eve} is
given by $V^{(0)} = (v_{00}, v_{10},v_{20},v_{30})$, its overall normalization
is such that, with our conventions,
\be
|V^{(0)}|^2 \, =\, v_{00}^2 \,-\, v_{10}^2 \,-\,v_{20}^2 \,-\,v_{30}^2 \,=\,1
\label{eq:norm0}
\ee
whereas, for the other three eigenvectors $V^{(i)}$, we must have
\be
|V^{(i)}|^2 \, =\, v_{0i}^2 \,-\, v_{1i}^2 \,-\,v_{2i}^2 \,-\,v_{3i}^2 \,=\,-1.
\label{eq:normi}
\ee
\item We now build a rotation matrix $O$, with the eigenvectors
$V^{(a)}$ serving as its columns. Which means, with the coefficients $v$ used in
eqs.~\eqref{eq:norm0} and~\eqref{eq:normi}, $O_{ab} = v_{ab}$. This matrix $O$
satisfies
\be
O^{-1}\,\Lambda\,O \, = \, \mbox{diag}(\Lambda_0\,,\,\Lambda_1\,,\,
\Lambda_2\,,\,\Lambda_3).
\ee
With this $4\times 4$ matrix, we can obtain 4-vectors $M^\mu$ and $r^\mu$, 
evaluated at the vacuum with neutral vevs $v_1/\sqrt{2}$ and
$v_2/\sqrt{2}$, in the basis where $\Lambda_{\mu\nu}$ is diagonal. In other
words, the quantities $\hat{M}^a$ and $\hat{r}^a$ are obtained by
\be
\left(\begin{array}{c}
\hat{M}^0 \\ \hat{M}^1 \\ 0 \\ \hat{M}^3
\end{array}\right)\,=\,
O^T\,
\left(\begin{array}{c}
-\frac{1}{2}(m_{11}^2 + m_{22}^2) \\ -\mbox{Re}\left( m_{12}^2 \right) \\
0 \\ \frac{1}{2}(m_{11}^2 - m_{22}^2)
\end{array}\right)\;\;\; , \;\;\;
\left(\begin{array}{c}
\hat{r}^0 \\ \hat{r}^1 \\ 0 \\ \hat{r}^3
\end{array}\right)\,=\,
O^T\,
\left(\begin{array}{c}
\frac{1}{2}(v_1^2 + v_2^2) \\ v_1\,v_2 \\
0 \\ \frac{1}{2}(v_1^2 - v_2^2)
\end{array}\right).
\label{eq:mhat}
\ee
\end{itemize}
And thus $\hat{M}_0 = \hat{M}^0$, $\hat{M}_i = -\hat{M}^i$, etc.
Since we began with the CP-conserving potential of eqs.~\eqref{eq:Mmu}
and~\eqref{eq:Lambda}, we are guaranteed to obtain
$\hat{M}_2 = \hat{r}_2 = 0$~\footnote{Remember, though, that CP breaking
vacua, which would have $\hat{r}_2 \neq 0$, are excluded from the start, due
to the existence of, at least, one normal minimum.}.
Now in possession of the values of the eigenvalues $\Lambda_0$, $\Lambda_i$; of the
rotated quadratic coefficients $\hat{M}_0$, $\hat{M}_i$; and of the rotated vevs
$\hat{r}_0$, $\hat{r}_i$, the necessary conditions for existence of two neutral minima are
very simple to write:
\bea
 & & \mbox{If } \hat{M}_0 \,>\,0\;\mbox{and}  \;\sqrt[3]{x^2} + \sqrt[3]{y^2}
\leq  1, \;\mbox{with} \no
 & & \no
 & & x = \frac{\hat{M}_1\,(\Lambda_0 - \Lambda_3)}{\hat{M}_0\,(\Lambda_3 - \Lambda_1)}
 \;\;,\;\;
 y  = \frac{\hat{M}_3\,(\Lambda_0 - \Lambda_1)}{\hat{M}_0\,(\Lambda_3 - \Lambda_1)}, \no
  & & \no
   & & \mbox{{\em then} the potential can have two neutral minima.}
\label{eq:cond_two}
\eea
We emphasize that these are {\em necessary} conditions for existence of two neutral
minima (see appendix~\ref{ap:min}) - although they are {\em necessary and sufficient} conditions for the existence
of four normal stationary points. Remarkably, though, we have a necessary and
sufficient condition to verify the global nature of our minimum -
to know whether our $\{v_1 , v_2\}$ vacuum is the global minimum of the
potential, we need only do the following:
\bea
 & & \mbox{Let us define a discriminant $D$, given by} \no
 & & \no
 & & D\,= \, \hat{M}_1\,\hat{M}_3\,\hat{r}_1\,\hat{r}_3; \no
 & & \no
 & & \mbox{{\em Our vacuum is the global minimum of the potential if and only if}}\; D> 0.
\label{eq:cond_disc}
\eea
It is simple to verify that this procedure leads to the
conditions laid out for the softly broken $Z_2$ model in section~\ref{sec:Z2}.
Unfortunately, the diagonalization procedure explained above renders analytical
expressions for the bounds unviable, in the case of the most general CP-conserving
potential. But the ``recipe" we provided in this section is quite easy to implement
in a numerical study.

\section{Panic vacuum bounds: a demonstration}
\label{sec:dem}

We will now demonstrate how the conditions for the panic vacua presented in the previous
section are obtained. We are assuming scalar potentials which are stable in a strong sense
(in the terminology of \cite{heidelberg}): that is, $\Lambda_{\mu\nu} r^\mu r^\nu > 0$
everywhere on and in the forward lightcone (apart from the apex).
Potentials stable in a weak sense cannot have two-minima configurations~\cite{ivanovPRE},
so we do not consider them.

It was shown in~\cite{Ivanov:2006yq,Ivanov:2007de} that, for potentials stable in
a strong sense, $\Lambda_{\mu\nu}$ can be always diagonalized by an $SO(1,3)$ transformation.
This corresponds to the ``recipe" we presented in the previous section. We therefore assume
that we have performed that diagonalization, and the tensor $\Lambda_{\mu\nu}$ is written as
\be
\Lambda_{\mu\nu} = \mathrm{diag}(\Lambda_0,\, -\Lambda_1,\, -\Lambda_2,\,-\Lambda_3)\,,
\ee
with the $\Lambda_a$ coefficients satisfying the conditions of
eqs.~\eqref{eq:bfb1}--~\eqref{eq:bfb3} so that the potential is bounded from below.
Likewise, the 4-vector of dimension-two coefficients is now given by $\hat{M}_\mu = (\hat{M}_0 \,,\,
\hat{M}_1 \,,\, 0 \,,\, \hat{M}_3)$, calculated using eq.~\eqref{eq:mhat}. Although this diagonalization
does not preserve the form of the scalars' kinetic terms, for purposes of determining the
extrema of the potential that is not problematic. With the potential written in terms of
these parameters, the minimization problem is reduced to search for points lying on the surfaces of
the forward lightcone, $LC^+$, which minimize the potential.

We remind the reader that we are interested only in neutral minima, and in particular in the
possibility of two neutral CP conserving minima coexisting. Thus, we will not worry about the
possibility of CP, or CB, minima, as they cannot exist when a normal minimum does.
For instance, spontaneous CP violation could only occur if
$\Lambda_2 > \Lambda_1, \Lambda_3$ and the $\hat{M}_a$
obeyed the following condition:
\be
{\hat{M}_1^2 \over (\Lambda_2 - \Lambda_1)^2} + {\hat{M}_3^2 \over (\Lambda_2 - \Lambda_3)^2} <
{\hat{M}_0^2 \over (\Lambda_0 - \Lambda_2)^2}\,.
\label{eq:sCPV}
\ee
Therefore, our parameters have to be such that either $\Lambda_2$ is not larger than $\Lambda_1$ and
$\Lambda_3$ or the condition of eq.~\eqref{eq:sCPV} is not verified.
But in fact, it is simple to show that requiring the pseudoscalar squared mass $m^2_A$ to
be positive (a necessary condition for a normal minimum to occur) implies that $\Lambda_2$ is
not the largest of the $\Lambda_i$. We therefore assume that we are in that situation.

Let us then go through the demonstration of the several conditions necessary for the existence
of two normal minima - eqs.~\eqref{eq:cond_two} - and the construction of the discriminants which
establish how panic vacua occur - eq.~\eqref{eq:cond_disc}.

\subsection{Condition $\hat{M}_0 > 0$}
\label{sec:M0}

Here we show that, if $\hat{M}_0< 0$ in the $\Lambda_{\mu\nu}$-diagonal basis,
then the potential has only one non-zero stationary point, which must be the global minimum.
The method we use is essentially the same as in~\cite{heidelberg} and~\cite{Ivanov:2007de}.

Finding neutral extrema of the potential~\eqref{eq:potr}
- that is, with values of $r^\mu$ restricted to the surface of lightcone,
$r^\mu r_\mu = 0$ - benefits from using a Lagrange multiplier $\zeta$. We
introduce an auxiliary potential $\bar{V} \,=\, V \,-\, \zeta\,r^\mu r_\mu/2$,
and minimize it, with respect to both the $r^\mu$ and $\zeta$. The
minimization conditions thus become
\be
\Lambda_{\mu\nu} r^\nu - M_\mu = \zeta r_\mu\,.
\ee
Using the explicit coefficients of the potential in the $\Lambda_{\mu\nu}$-diagonal frame ,
the minimization conditions become
\be
(\Lambda_0 - \zeta) \hat{r}_0 = \hat{M}_0\,, \quad (\Lambda_i - \zeta) \hat{r}_i = \hat{M}_i\,.
\label{eq:system}
\ee
Notice that, since the potential is CP conserving and no CP spontaneous breaking is being
considered, $\hat{M}_2 = 0$ and $\hat{r}_2 = 0$, always. Thus in these equations $i = 1, 3$.
This system has therefore three independent variables: $\zeta$ and two $\hat{r}_i$ -
the value of $\hat{r}_0$ is then expressed as the positive square root of $\sum_i \hat{r}_i^2$.

Now, since $\hat{r}_0$ is necessarily positive~\footnote{Taking $\hat{r}_0 >0$ means we are excluding the
trivial solution, all $\hat{r}_\mu = 0$.}, the first equation in~\eqref{eq:system} implies that,
if $\hat{M}_0 < 0$, the solution is found for a value of the Lagrange multiplier $\zeta > \Lambda_0$.
Notice, also, that the condition $\hat{r}_0^2 = \sum \hat{r}_i^2$ can be rewritten as
\be
\sum_{i=1,3} \left({\zeta - \Lambda_0 \over \zeta - \Lambda_i}\right)^2 \cdot \left({\hat{M}_i \over \hat{M}_0}\right)^2 = 1\,.
\label{eq:zeta}
\ee
This is a single algebraic equation of fourth order in $\zeta$.
Even without solving it, we can extract much information
from it .

Let us vary the value of $\zeta$ from $\Lambda_0$ to infinity. We see that the variables
$(\zeta -\Lambda_0)/(\zeta - \Lambda_i)$ increase, in a monotonous manner,
from zero to one. The expression on the left of~\eqref{eq:zeta} is therefore a monotonous
function of $\zeta$, and it grows from zero to a maximum equal to $(\sum \hat{M}_i^2)/\hat{M}_0^2$.

We can therefore conclude that:
\begin{itemize}
\item
If $\hat{M}_0 < 0$ and $\hat{M}_\mu \hat{M}^\mu \ge 0$, then $(\sum \hat{M}_i^2)/\hat{M}_0^2 \le 1$, and the equation~\eqref{eq:zeta} has no solution in the region $\zeta > \Lambda_0$.
Thus, the potential has no non-trivial extremum.

The only extremum - the global minimum - lies at $r^\mu = 0$, and no electroweak breaking occurs. This
situation is therefore physically uninteresting.
\item
If $\hat{M}_0 < 0$ and $\hat{M}_\mu \hat{M}^\mu < 0$, then $(\sum \hat{M}_i^2)/\hat{M}_0^2 > 1$, and the equation~\eqref{eq:zeta} has only one solution at $\zeta > \Lambda_0$. This is the only non-trivial stationary point of the potential, and it must be the global minimum, as it corresponds to $V < 0$.

Thus no two minima can occur in this situation.
\end{itemize}
Therefore, $\hat{M}_0 > 0$ is a necessary condition for the 2HDM potential to have two minima.
Since  $r_0 >0$ per definition, eq.~\eqref{eq:system} implies that
$\zeta < \Lambda_0$.

\subsection{The astroid condition}
\label{sec:ast}

Here we show that, for the 2HDM scalar potential to have two normal minima, the values of
the potential's parameters must be such that we are inside a region of space limited by the
astroid curve defined in eq.~\eqref{eq:cond_disc}. We will use the geometric approach to counting
solutions of the minimization equations~\eqref{eq:zeta} developed in~\cite{Ivanov:2007de}. Please
remember that we are only interested in extrema with vevs without any relative phase,
and thus $r_2 = 0$. Since the existence of a normal minimum precludes a CP-breaking one,
this means that a more generic analysis would discover a greater number of saddle points.
But restricting ourselves to the $r_2 = 0$ case has no impact on the counting of possible
normal minima. The analysis has a subtlety related to the ordering of the eigenvalues
$\Lambda_i$. Let us start with the case where $\Lambda_1\,>\,\Lambda_3$~\footnote{And also
$\Lambda_1\,>\,\Lambda_2$, since we are precluding the possibility of a CP-violating minimum.}.

Eq.~\eqref{eq:zeta} can be seen as defining an ellipse. In fact, if we define the variables
$m_1 = \hat{M}_1 /\hat{M}_0$ and $m_3 = \hat{M}_3 /\hat{M}_0$, the semiaxes of the ellipse will
depend on $\zeta$ and will be given by
\be
a_1 = |\Lambda_1 - \zeta|/(\Lambda_0 - \zeta)\;\;\;, \;\;\;
a_3 = |\Lambda_3 - \zeta|/(\Lambda_0 - \zeta)\;.
\label{eq:axes}
\ee
In terms of these new variables eq.~\eqref{eq:zeta} thus becomes
\be
\frac{m_1^2}{a_1^2} \;+\;\frac{m_3^2}{a_3^2} \; = \; 1,
\ee
which is clearly the equation describing an ellipse in the $(m_1\,,\,m_3)$ plane.
Let us consider the family of ellipses which we obtain when we take all values of the
Lagrange multiplier, $-\infty < \zeta < \Lambda_0$, and count how many times this family
of ellipses crosses a specific point $(a\,,\,b)$ in the
$m_1\times m_3$ plane. When an ellipse passes by that point,
eq.~\eqref{eq:zeta} has a solution, which means that the potential has an extremum.
Then, the number of times the ellipses pass over the point $(a\,,\,b)$ will give us the
number of non-trivial extrema of the potential. Also, it has been shown (for instance,
in~\cite{heidelberg}) that the larger $\zeta$ is, the smaller the value of the potential.
\begin{figure}[ht]
\centerline{
\includegraphics[height=8cm,angle=0]{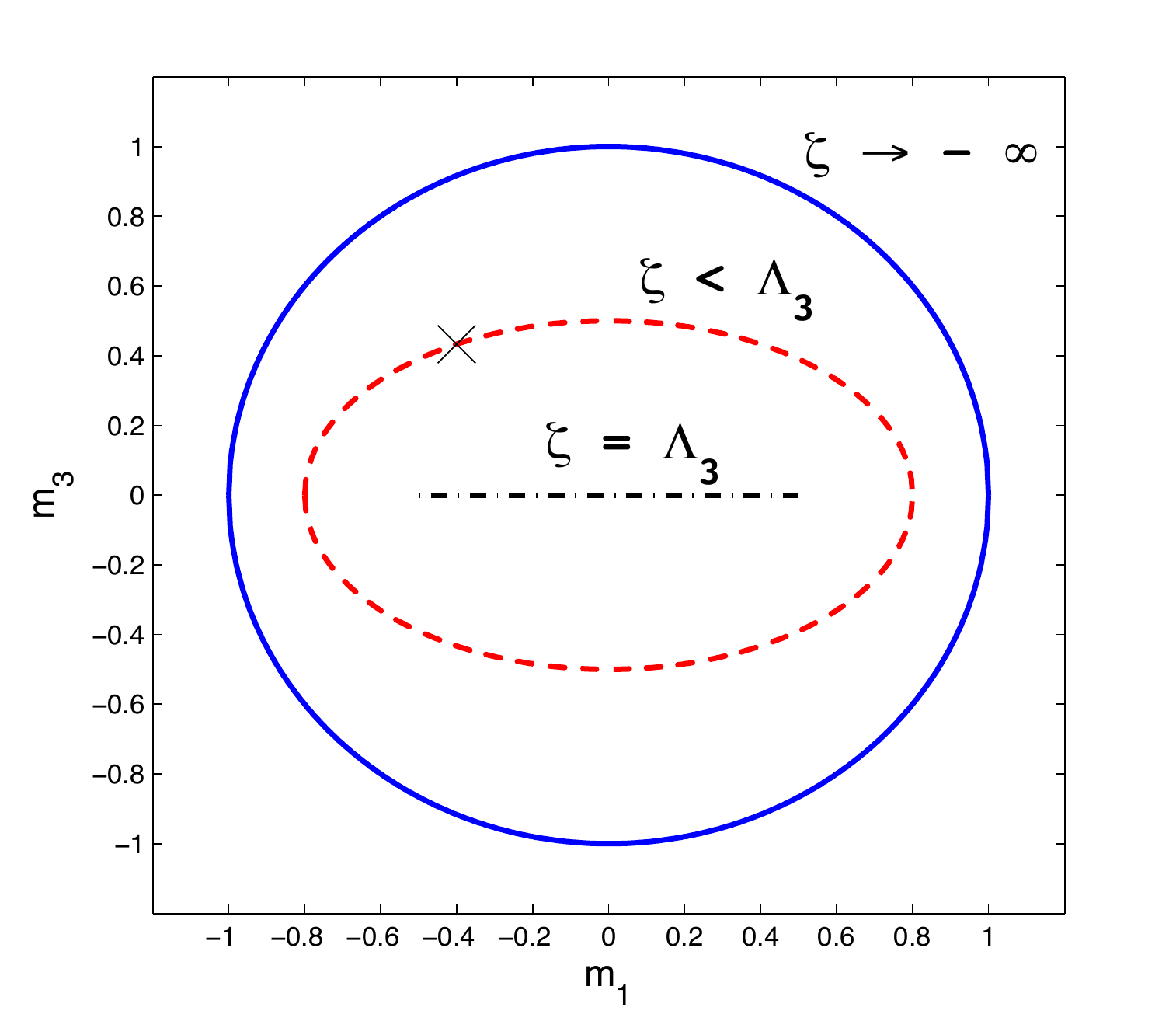}
\includegraphics[height=8cm,angle=0]{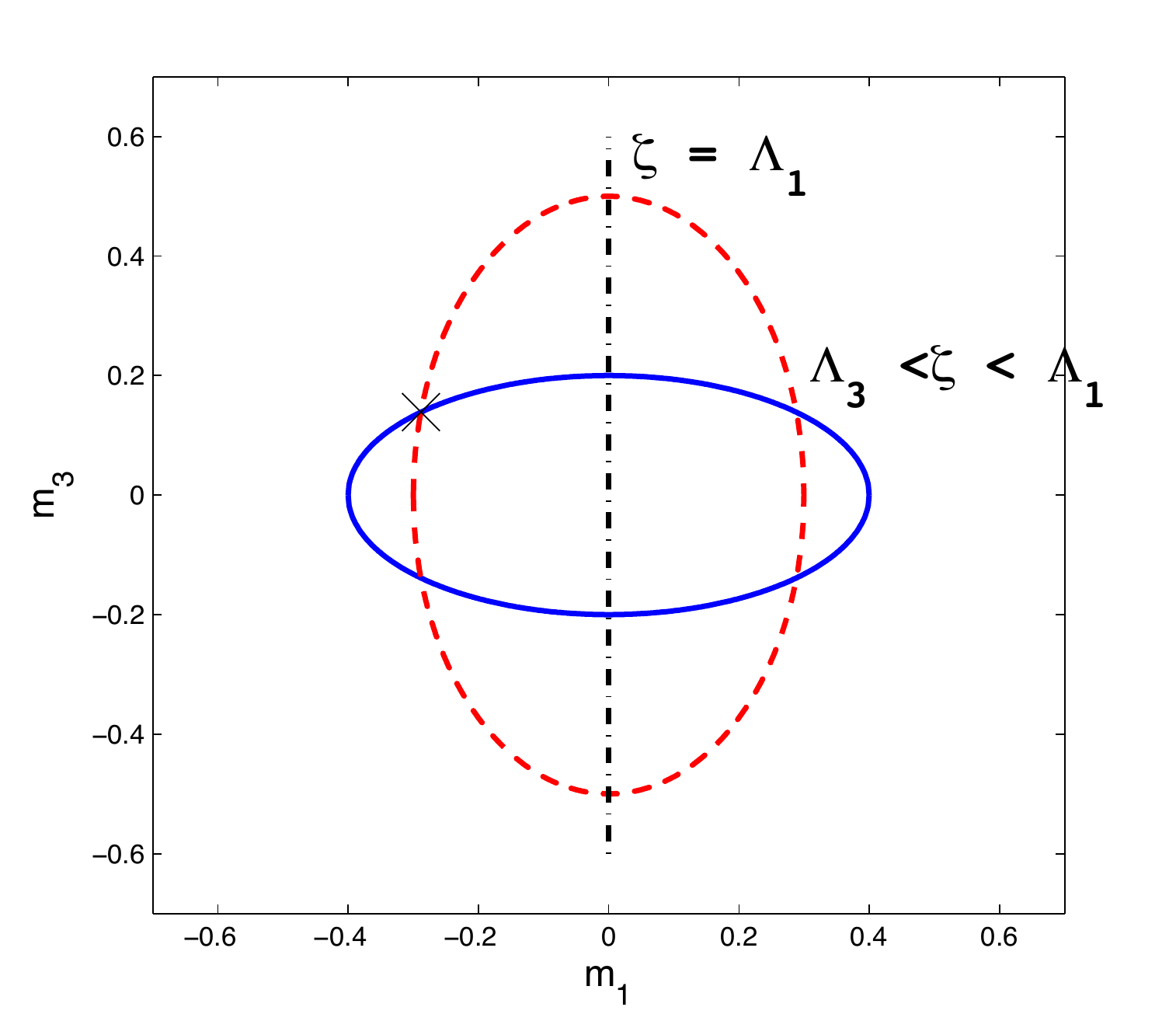}}
\caption{Example of the geometric evolution of the ellipse in eq.~\eqref{eq:zeta} with the value of
the Lagrange multiplier. On the left, the range $-\infty < \zeta \leq \Lambda_3$;
on the right, $\Lambda_3 < \zeta \leq \Lambda_1$. }
\label{fig:eli1}
\end{figure}

So, as $\zeta$ changes from $-\infty$ to $\Lambda_0$, we have:
\begin{itemize}
\item
In the limit $\zeta \to -\infty$, the ellipse is just the unit circle. This occurs because the semiaxes
$a_1$ and $a_3$ tend to one in this limit (solid blue line in fig.~\ref{fig:eli1}, left).
\item
As $\zeta$ increases, the ellipse shrinks. It shrinks most quickly along the direction
of the smallest semiaxis - and since $\Lambda_1 > \Lambda_3$, then, according to eq.~\eqref{eq:axes},
this means that the ellipse is contracting faster along the axis $m_3$ (dashed red line
in fig.~\ref{fig:eli1}, left).
\item
As $\zeta$ increases further, the ellipse on the $(m_1,m_3)$-plane shrinks even more. At
$\zeta=\Lambda_3$ the ellipse collapses to a line segment (dot-dashed black line in
fig.~\ref{fig:eli1}, left):
\be
m_3 = 0\,,\quad |m_1| \le m_1^* = {\Lambda_1 - \Lambda_3 \over \Lambda_0 - \Lambda_3}\,.
\ee
\item
Notice that over the interval $-\infty < \zeta < \Lambda_3$ {\em the ellipses sweep exactly once} all points inside the unit circle. For instance, there is only one ellipse which passes by the point $(a\,,\,b)$ marked with an ``X" in fig.~\ref{fig:eli1}, left.
\end{itemize}
A single crossing means that in this region there is only one non-trivial solution of the minimization equations
for this region of values of $\zeta$.

For $\Lambda_3 < \zeta < \Lambda_1$ the situation is different:
\begin{itemize}
\item
For $\zeta > \Lambda_3$ the line segment again becomes an ellipse.
When $\zeta$ increase further, $\Lambda_3 < \zeta < \Lambda_1$, the ellipse shrinks along the $m_1$ axis and grows along $m_3$ axis (solid blue and dashed red lines in fig.~\ref{fig:eli1}, right).
\item
At $\zeta=\Lambda_1$ the ellipses collapse to another line segment (dot-dashed black line in
fig.~\ref{fig:eli1}, right):
\be
m_1 = 0\,,\quad |m_3| \le m_3^* = {\Lambda_1 - \Lambda_3 \over \Lambda_0 - \Lambda_1} > m_1^*\,.
\ee
\item
During this evolution, the ellipses sweep a certain region in the $(m_1\,,\,m_3)$-plane, and {\em each point inside this region is crossed exactly twice} (see, for instance, the point marked with an ``X" in fig.~\ref{fig:eli1}, right).
\item
It is possible to show (see Appendix~\ref{ap:min}) that, of these two crossings, one is a saddle point and,
{\em if one imposes further conditions on the parameters}, the other can be a minimum.
\end{itemize}
Finally, when $\Lambda_1 < \zeta < \Lambda_0$, the ellipse grows infinitely (the semiaxes $a_1$ and $a_3$
tend to infinity when $\zeta \rightarrow \Lambda_0$), and the entire  $(m_1\,,\,m_3)$-plane is covered. But
{\em each point of the plane is crossed by an ellipse one single time}. And since this last crossing
corresponds to the largest value of $\zeta$ to yield an extremum, {\em it corresponds to the global
minimum of the potential}. Please notice: regardless of whether the previous
stationary points exist or not, this one is always present, and it is always the global minimum of
the potential.

We see, therefore, that we have the possibility of two minima - one may lie in the region where $\Lambda_3 < \zeta < \Lambda_1$ (if we impose further constraints on the parameters of the potential), another (guaranteed)
for values of $\zeta$ such that $\Lambda_1 < \zeta < \Lambda_0$. Had we considered the
case where $\Lambda_3\,>\,\Lambda_1$, we would have similar conclusions: the figures obtained for that
situation would be analogous to those in fig.~\ref{fig:eli1}, but rotated by 90$^o$, and both minima
would be found for values of the Lagrange multiplier in the ranges $\Lambda_1 < \zeta < \Lambda_3$ and
 $\Lambda_3 < \zeta < \Lambda_0$. Notice that all we have been able to ensure, with the two
conditions~\eqref{eq:M0} and~\eqref{eq:cond_disc}, is the existence of four normal
stationary points, with one of them being certainly the global minimum of the potential. To
make certain that one of the remaining stationary points is the second minimum we would need
further conditions.

Two neutral minima are therefore possible if, in the $(m_1\,,\,m_3)$-plane, the values of the parameters of
the potential are such that we are in a point inside the region of space covered by intersecting ellipses such
as the two shown in fig.~\ref{fig:eli1}, right. {\em The region covered by all of those ellipses is delimited
by the astroid curve given in eq.~\eqref{eq:cond_two}}. To prove this, we apply the standard method of
finding the envelope of a family of curves -  the family of ellipses in question is generically described
by an equation of the form $F(m_1\,,\,m_3\,,\,\zeta) \,=\,0$, and we also need to consider the tangent to
these ellipses at each point, determined by $\partial F/ \partial \zeta\,=\,0$. Explicitly, we need to
solve
\bea
F(m_1\,,\,m_3\,,\,\zeta) & = & \frac{m_1^2}{(\Lambda_1 - \zeta)^2}\;+\;
\frac{m_3^2}{(\Lambda_3 - \zeta)^2}\;-\;\frac{1}{(\Lambda_0 - \zeta)^2}\;=\;0 \label{eq:env1}\\
\frac{1}{3}\,\frac{\partial F}{\partial \zeta} & = & \frac{m_1^2}{(\Lambda_1 - \zeta)^3}\;+\;
\frac{m_3^2}{(\Lambda_3 - \zeta)^3}\;-\;\frac{1}{(\Lambda_0 - \zeta)^3}\;=\;0\;.
\label{eq:env2}
\eea
To solve these equations we define
\be
\cos\varphi\,=\,\frac{\Lambda_0 - \zeta}{\Lambda_1 - \zeta}\,m_1\;\;\; ,\;\;\;
\sin\varphi\,=\,\frac{\Lambda_0 - \zeta}{\Lambda_3 - \zeta}\,m_3\;,
\label{eq:cs}
\ee
so that eq.~\eqref{eq:env1} is automatically satisfied, and eq.~\eqref{eq:env2} becomes
\be
\frac{\Lambda_0 - \Lambda_1}{\Lambda_1 - \zeta}\,\cos^2\varphi\;+\;
\frac{\Lambda_3 - \Lambda_0}{\Lambda_3 - \zeta}\,\sin^2\varphi\;=\;0.
\ee
From these two equations we determine the value of $\zeta$ which, substituted in eqs.~\eqref{eq:cs},
yields
\be
\cos^3\varphi\,=\,\frac{\Lambda_0 - \Lambda_3}{\Lambda_1 - \Lambda_3}\,m_1\;\;\; ,\;\;\;
\sin^3\varphi\,=\,\frac{\Lambda_0 - \Lambda_1}{\Lambda_3 - \Lambda_1}\,m_3\;.
\ee
Recalling that $m_i = \hat{M}_i/\hat{M}_0$ and substituting these expressions in $\cos^2\varphi +
\sin^2\varphi = 1$, we obtain the astroid curve which appears in our criterion for existence
of two minima, eq.~\eqref{eq:cond_two}~\footnote{Notice that a sign difference in one of the denominators
is irrelevant, due to the squares involved.}, {\em Q.E.D}.

\subsection{Conditions for panic vacua}

The necessary and sufficient conditions for the existence of
panic vacua can be considered by themselves, circumventing the need to verify whether or not
the potential has two minima.

The demonstration is extremely simple. Let us begin with the case
$\Lambda_1 > \{\Lambda_2\,,\,\Lambda_3\}$. As we have seen in the previous section, in
this situation the several possible stationary points obey the following relations:
\begin{itemize}
\item
The global minimum occurs for a value of the Lagrange multiplier, $\zeta_G$, such that
$\zeta_G > \Lambda_1 > \Lambda_3$.
\item
If another, local, minimum exists, it can only occur for a given value of the Lagrange multiplier,
$\zeta_L$, such that $\Lambda_3 < \zeta_L < \Lambda_1$.
\end{itemize}

And this is all the information we require. Recalling the minimization conditions
of the potential, written in terms of the Lagrange multiplier (eqs.~\eqref{eq:system}),
we define the discriminants,
\be
D_1 = - \hat{r}_1 \hat{M}_1\,, \quad D_3 = - \hat{r}_3 \hat{M}_3\,, \quad D = D_1 D_3\,.
\ee
Given the minimization conditions, eqs.~\eqref{eq:system}, we can write
\bea
D_1 & = & -\hat{r}_1 \hat{M}_1 \; = \; (\zeta - \Lambda_1)\,\hat{r}_1^2\;\;\; , \no
D_3 & = & -\hat{r}_3 \hat{M}_3 \; = \; (\zeta - \Lambda_3)\,\hat{r}_3^2\;\;\;.
\eea
These discriminants can be computed for any minimum, {\em i.e.} for any given
value of $\zeta$. Then, we see that:
\begin{itemize}
\item
In the global minimum $\zeta = \zeta_G$.
\item
Given that $\zeta_G > \Lambda_1 > \Lambda_3 $, we will have $D_1 >0$ and $D_3 >0$.
\item
{\em \bf{Thus, at the global minimum, $D = D_1 D_3 > 0$}}.
\item
If the second, local, minimum exists, it occurs for $\zeta = \zeta_L$.
\item
Since $\Lambda_3 < \zeta_L < \Lambda_1$, we will necessarily have
$D_1 < 0$ and $D_3 >0$.
\item
{\em \bf{Thus, at the local minimum, $D = D_1 D_3 < 0$}}.
\end{itemize}
And so we see that the sign of $D_1$ discriminates between the local and the global
minima, while the sign of $D_3$ does not. If it happens that the potential has only one minimum,
it will correspond to the case $D_i > 0$~\footnote{Also, in the case where $\hat{M}_0 < 0$
and there is a single minimum with $\zeta_G > \Lambda_0$, discussed in section~\ref{sec:M0}, both
$D_1$ and $D_3$ are guaranteed to be positive, given that boundedness from below implies $\Lambda_0 > \{\Lambda_1\,,\, \Lambda_3\}$.}.

Suppose we now have $\Lambda_1 > \Lambda_3$. The demonstration for this case is analogous to
the one we have just given, with the following differences: the global minimum is now at
$\zeta = \zeta_G > \Lambda_3 > \Lambda_1$; the local minimum, if it exists, corresponds to a
Lagrange multiplier $\Lambda_1 < \zeta_L < \Lambda_3$; at the global minimum we will have $D_1 > 0$
and $D_3 > 0$, at the local one $D_1 > 0$ and $D_3 <0$. Thus, in this case, the sign of $D_3$ does
discriminate between the local and the global minima, but the sign of $D_1$ does not. In any case,
$D = D_1 D_3$ is positive at the global minimum and negative at the local one.

In conclusion, the product of $D_1$ and $D_3$ is a quantity able to discriminate between the two normal minima:
if we calculate it at a given minimum and find $ D\, =\, D_1 D_3\,>\,0$, that minimum is the global
minimum of the potential; if $D\,<\,0$, the minimum is local. Thus the conditions shown
in~\eqref{eq:cond_disc} are proven.

\section{Lifetime of the metastable vacuum}
\label{sec:life}

If particle physics is described by the 2HDM and $D\,<\,0$, we are in a metastable minimum,
and there is the possibility of tunneling to the true vacuum of the model.
It can be argued, though, that the existence of these
panic vacua is not sufficient reason to exclude the parameters of the potential which produce
them. In fact, if the tunneling time to the true vacuum is superior to the current age of the
universe, the existence of a panic vacuum is completely acceptable. If, on the other hand, the
tunneling time is inferior to the age of the universe,
that region of parameter space ought to be excluded since it predicts a vacuum which would
have already decayed, contrary to current experimental evidence.

As such, it is interesting to try to estimate, for panic vacua, the decay width density,
$\Gamma/V$, for a universe in such a metastable state. This is an application of the classic
calculation by Coleman~\cite{cole} for a potential with a single scalar field, very well
discussed in the book by Rubakov~\cite{ruba}. Let us consider a
potential with two minima, such that their difference in depths is given by $\epsilon$;
there is a maximum between both minima, with height $\delta$ relative to the local minimum.
The value of the potential at the local minimum is taken to be zero. Then, one finds that
\be
\frac{\Gamma}{V}\;=\;A\,e^{-B}\;,
\label{eq:wid}
\ee
where $A$ is a small prefactor, in comparison with the exponential of $-B$. The calculation of $B$,
even for the simple case of a single scalar field,
is very involved and requires a series of assumptions and approximations. Namely, it is assumed
that: the height of the barrier $\delta$ and the relative depth of the
minima $\epsilon$ satisfy $\delta/\epsilon \,\gg\,0.06$~\cite{cole}; that the only path linking both minima passes through a
maximum of height $\delta$ (which is a trivial assumption for a potential depending on a
single field); and, while studying the expansion of the bubble corresponding to the true
vacuum as the universe tunnels to it, one considers the so-called ``thin wall" approximation,
considering that the border between the regions of the universe lying in different vacua,
as the bubble expands, has no thickness~\cite{colebook}. In this case,
\be
B \; =\; \frac{2^{11} \pi^2}{3 \lambda}\,\left(\frac{\delta}{\epsilon}\right)^3\;,
\label{eq:life}
\ee
where $\lambda$ is the quartic coupling of the scalar potential. To have a rough estimate of what
happens in the 2HDM case, we take $\lambda = \mbox{max}|\lambda_k|$, with $k = 1, \ldots 5$. But
notice that, since in our potential we may have two minima and a multitude of saddle points and
maxima, $\delta$ is not necessarily the height of a maximum relative to the false minimum, but
rather taken to be the height of the lowest stationary point - saddle point or maximum - relative to it.

Considering the numeric factors in eq.~\eqref{eq:life}, and the fact that $\lambda$ is limited in
size by unitarity considerations, if the ratio $\delta/\epsilon$ is larger
than about 1, the quantity $B$ will be quite sizeable, and as such the decay width of
eq.~\eqref{eq:wid} becomes extremely small - which means that the lifetime of the false
vacuum, the inverse of $\Gamma$, becomes extremely large, and no tunneling occurs during the
lifetime of the universe. On the contrary, sets of parameters which predict $\delta/\epsilon$
smaller than about 1 ought to be excluded from the model's parameter space, since they produce
a tunneling time smaller than the age of the universe.

We have computed this estimate for the lifetime for all our panic vacua. Using the
equation~\eqref{eq:zeta}, it is easy to discover all possible
normal stationary points, thus determining both $\epsilon$ and $\delta$. We find that
the vast majority of the panic vacua do {\em not} lead to vacuum lifetimes
larger than the age of the universe. Quite the contrary, they correspond to very small
values of $\delta/\epsilon$ - either in potentials with minima of extremely different
depths, or barriers of small height - which would mean that these vacua would have decayed long ago.
If one believes this estimate, then, they must definitely be excluded. A very small percentage
of points - less than 3\% of the total of panic vacua found - seems, however, to have
a false vacuum lifetime larger than the age of the universe, and as such ought to be retained.
Our conclusions remain essentially unchanged, since the
percentage of ``safe panic vacua" is indeed very small.

We must however point out several shortcomings in this lifetime estimate. In fact, the thin-wall
approximation which leads to obtaining the estimate of eq.~\eqref{eq:life} breaks down if
$\delta/\epsilon \leq 0.1$ - and for most of our panic vacua $\delta/\epsilon$ is well below this
value. As such, this estimate is problematic at best. Also, and perhaps even more serious,
the estimates shown herein, which are widely used in the literature, are obtained in a model
with a single scalar field. In the 2HDM, however, even if we use the gauge freedom to exclude
the would-be Goldstone bosons, we are left with a potential with {\em five real scalar fields}.
Now, to calculate the decay width density, we have to determine the bounce trajectory between
both minima. This corresponds to a classical analysis of the scalar action in Euclidean space-time.
In a scalar potential which is five-dimensional in its field content, the determination of
the bounce is impossible to do analytically. Instead, we assumed that this trajectory corresponds,
as we already explained, to the one that goes over the lowest intermediate saddle point,
following the path of steepest descent. Although this seems like a reasonable approximation, it
is known~\cite{ruba} that in multidimensional problems the bounce trajectory sometimes avoids the saddle
point - in simpler terms, there is an easier path between both minima which avoids
the intermediate saddle point. Thus, the fact that we found points which seem to be safe for
tunneling is not too reassuring - there may well be another path between both minima which
leads to much smaller lifetimes of the panic vacua.

The remarkable thing, though, is that we can, in large measure, sidestep any cosmological considerations
and simply look at what the LHC data tells us about the nature of the 2HDM vacuum. As we see for model I
 all panic vacua are excluded at $2\sigma$, {\em regardless of what their estimated
lifetime might be}. If however one chooses to believe in the traditional lifetime estimate for the
false vacua, one will still find that most of the parameter space where panic vacua occur should
indeed be excluded from phenomenological analysis, given that the lifetimes of those vacua would
be far inferior to the age of the universe.

\section{Conclusions}

The 2HDM scalar potential has a very rich vacuum structure, and the possibility of the coexistence
of two normal minima in the tree-level is well-established. In this work we have derived the conditions that
the parameters of the potential have to obey so that two normal minima may exist. We have also
shown how one can build simple discriminants which allow us to conclude whether or not a given vacuum is
the global minimum of the potential. These discriminants take on a particularly simple form for the most
used 2HDM potential, the model with a softly broken $Z_2$ symmetry, with explicit CP conservation. In
this case, the vacuum of the model is global if and only if
\be
D \,=\, m^2_{12} (m^2_{11} - k^2 m^2_{22}) (\tan\beta - k)\,>\,0\,,
\ee
with $k = \sqrt[4]{\lambda_1/\lambda_2}$.

We have performed a thorough scan of the parameters of the softly broken $Z_2$ model and shown
that the occurrence of two normal minima is not confined to a non-interesting corner of parameter space:
two normal minima occur very often in this model, if one does a blind choice of parameters. Further,
we have seen that current LHC data disfavour at the $2\sigma$ level the possibility that,
for Higgs Yukawa interactions of Model I,
the minimum of the model which has $v = 246$ GeV and $m_h = 125$ GeV is {\em not} the global minimum
of the potential. Thus, and even before any cosmological
considerations, current particle physics experimental data already tells us a great deal about the
nature of the 2HDM vacuum. However, for Model II, current LHC data cannot exclude the possibility
that the model's vacuum is metastable, even at the $2\sigma$ level. This would mean that the model has
a deeper minimum, and the universe could, through tunneling, eventually reach that minimum.

Following the standard calculations on the subject, we performed an estimation of the tunneling times
involved in transition between vacua. To go beyond these simple estimates is extremely complex, and involves several
assumptions which might be debatable. Taken at face value, however, our estimate shows that the vast
majority of panic vacua we found would have lifetimes much smaller than the age of the universe.
Such a situation can definitely be excluded on anthropic grounds,
and the only panic vacuum points which we should really worry about are those
with lifetimes of the order of the age of the Universe.
But, as we discussed, our simple calculations might give an unreliable estimate
of the vacuum lifetime, and improving them is a very complicated task.
Thus, it is actually very difficult to establish which of the panic points can be truly worrying
and which can already be disregarded due to cosmological reasons.
The strong feature of our result is that we can exclude these points right now,
on the basis of the current LHC data, even without the need to improve the
reliability of the lifetime calculations. Another, possibly interesting, line of research,
considers the existence
of two simultaneous minima - but in the situation where no panic vacua occur, and the second minimum
lies above ``ours". What cosmological consequences might there be in such a situation? Could the universe
have rested, for a short while, in the upper minimum, before cooling down to ours? What observable
consequences would that have, if any?

The calculations performed in this paper were, of course, all undertaken at tree level. The importance
of radiative corrections to these results cannot, clearly, be dismissed. One may try to perform a
Renormalization Group improvement of these results, by using the 2HDM $\beta$-functions (see, for
instance,~\cite{Haber:1993an}) to run the values of the parameters of the potential with
the renormalization scale up to the mass scale typical of the problem at hand, to curtail the need
to compute the one-loop effective potential. However, please notice that both minima might have
vevs of very different magnitude, and as such each minima would have a very different ``typical
scale". It isn't clear, then, up to what scale one should run the parameters, since comparing the value
of the potential at both minima {\em at different scales} seems wrong. In fact, it has been argued,
in the context of charge breaking minima in supersymmetric models~\cite{Ferreira:2000hg,Ferreira:2001tk}
that the only correct procedure in the occurrence of two very distinct scales in two minima
one wishes to compare is to compute the one-loop effective potential at both minima, coupled with
an RG improvement of all parameters. Such a calculation is a gargantuan attempt. In this paper we
limit ourselves to drawing attention to the problem that is already present at tree level - the
existence of metastable, neutral vacua - and defer calculation of the impact of radiative corrections
to future work.

The panic vacua conditions presented in this work have a distinct practical advantage: they are
extremely simple to apply in a numerical study of the 2HDM. In fact, they are almost as simple
to apply as the usual bounded-from-below conditions, which are mandatory in any phenomenological
study, and which involve the numerical determination of the parameters of the potential. In
section~\ref{sec:pancp} we presented a method which can be applied to more general (but
still CP-conserving) versions of the 2HDM potential to obtain the corresponding panic
vacua conditions.

But even if one were to obtain precise estimates of the tunneling times to panic vacua, and found
them to be very superior to the age of the universe - which would remove any need to panic,
really - the conditions we present in this work would still be interesting, in that they would
allow us to determine, from particle physics experiments alone, a very interesting cosmological
information: the true nature of the universe's vacuum.

\begin{acknowledgments}
We thank Jo\~ao Paulo Silva for numerous important discussions and
participation in the early stages of this work.
The works of A.B., P.M.F. and R.S. are supported in part by the Portuguese
\textit{Funda\c{c}\~{a}o para a Ci\^{e}ncia e a Tecnologia} (FCT)
under contract PTDC/FIS/117951/2010, by FP7 Reintegration Grant, number PERG08-GA-2010-277025,
and by PEst-OE/FIS/UI0618/2011.
I.P.I. is thankful to CFTC, University of Lisbon, for their hospitality.
His work is supported by grants RFBR 11-02-00242-a,
RF President grant for scientific schools NSc-3802.2012.2, and the
Program of Department of Physics SC RAS and SB RAS "Studies of Higgs boson and exotic particles at LHC".
\end{acknowledgments}

\appendix

\section{Demonstration of panic vacuum bounds for the $Z_2$ model}
\label{ap:Z2}

The ``recipe" we give in section~\ref{sec:pancp} is completely general and it is a good exercise
for the reader to apply if to the case of the softly-broken $Z_2$ case and obtain the expressions
shown in section~\ref{sec:Z2}. However, in that case, there is an alternative manner to reach
the same results, and considerably simpler, and that is what we will now present.
We start with the Higgs potential for this model,
\begin{eqnarray}
V &=&
m_{11}^2 \Phi_1^\dagger \Phi_1 + m_{22}^2 \Phi_2^\dagger \Phi_2
- m_{12}^2 \left[  \Phi_1^\dagger \Phi_2 + \Phi_2^\dagger \Phi_1 \right]
+ {1 \over 2} \lambda_1 (\Phi_1^\dagger\Phi_1)^2
+ {1 \over 2} \lambda_2 (\Phi_2^\dagger\Phi_2)^2
\nonumber\\
&&+\,
\lambda_3 (\Phi_1^\dagger\Phi_1) (\Phi_2^\dagger\Phi_2)
+ \lambda_4 (\Phi_1^\dagger\Phi_2) (\Phi_2^\dagger\Phi_1)
+ {1 \over 2} \lambda_5 \left[(\Phi_1^\dagger\Phi_2)^2 + (\Phi_2^\dagger\Phi_1)^2\right]\,,
\label{VH1}
\end{eqnarray}
with all coefficients being real. The tensor $\Lambda_{\mu\nu}$ is not diagonal in this basis:
\be
\Lambda_{\mu\nu} = {1\over 2} \left(\begin{array}{cccc}
{\lambda_1 + \lambda_2 \over 2} + \lambda_3 & 0 & 0 & {\lambda_1 - \lambda_2 \over 2} \\
0 & \lambda_4 + \lambda_5& 0 & 0 \\
0 & 0 & \lambda_4 - \lambda_5& 0 \\
{\lambda_1 - \lambda_2 \over 2}& 0& 0 & {\lambda_1 + \lambda_2 \over 2} - \lambda_3
\end{array}\right)
\ee
In order to diagonalize it, we need to equilibrate $\lambda_1$ and $\lambda_2$. This can be achieved
by rescaling the doublets:
\be
\Phi_1 \to q \Phi_1\,, \quad \Phi_2 \to q^{-1} \Phi_2\,, \quad q = \left({\lambda_2 \over \lambda_1}\right)^{1/8}\,.
\ee
Upon this change, the potential becomes
\begin{eqnarray}
V &=&
m_{11}^2 q^2 \Phi_1^\dagger \Phi_1 + m_{22}^2 q^{-2}\Phi_2^\dagger \Phi_2
- m_{12}^2 \left[  \Phi_1^\dagger \Phi_2 + \Phi_2^\dagger \Phi_1 \right]
+ {1 \over 2} \sqrt{\lambda_1 \lambda_2}\left[(\Phi_1^\dagger\Phi_1)^2
+ (\Phi_2^\dagger\Phi_2)^2\right]
\nonumber\\
&&+\,
\lambda_3 (\Phi_1^\dagger\Phi_1) (\Phi_2^\dagger\Phi_2)
+ \lambda_4 (\Phi_1^\dagger\Phi_2) (\Phi_2^\dagger\Phi_1)
+ {1 \over 2} \lambda_5 \left[(\Phi_1^\dagger\Phi_2)^2 + (\Phi_2^\dagger\Phi_1)^2\right]\,,
\label{VH2}
\end{eqnarray}
In this basis, $\Lambda_{\mu\nu}$ is diagonal and has the following eigenvalues:
\be
\Lambda_0 = {1 \over 2}(\sqrt{\lambda_1 \lambda_2} + \lambda_3)\,,\quad
\Lambda_1 = - {1\over 2}(\lambda_4 + \lambda_5)\,,\quad
\Lambda_2 = - {1\over 2}(\lambda_4 - \lambda_5)\,,\quad
\Lambda_3 = {1 \over 2}(-\sqrt{\lambda_1 \lambda_2} + \lambda_3)\,.
\ee
Notice how the conditions of eqs.~\eqref{eq:bfb1}--~\eqref{eq:bfb3}
lead to the well-known bounded-from-below constraints.
In this basis, the components of vector $M_\mu$ take form:
\be
\hat{M}_0 = -{1 \over 2}(m_{11}^2 q^2 + m_{22}^2 q^{-2})\,,\quad \hat{M}_1 = m_{12}^2\,,\quad \hat{M}_2 = 0\,,\quad
\hat{M}_3 = {1 \over 2}(m_{22}^2 q^{-2} - m_{11}^2 q^2)\,.
\ee
If the neutral minimum point is parameterized as
\be
\langle \Phi_1^0\rangle = {v\over \sqrt{2}}\cos\beta \,,\quad
\langle \Phi_2^0\rangle = {v\over \sqrt{2}}\sin\beta\,,
\ee
then the components of $\langle r_\mu\rangle $ in the $\Lambda_{\mu\nu}$-diagonal basis
are
\be
\langle \hat{r}_1\rangle = -v^2 \sin\beta \cos\beta\,,\quad \langle \hat{r}_3\rangle = {v^2 \over 2}(q^2 \sin^2\beta -
q^{-2} \cos^2\beta)\,.
\ee
The two discriminants then become
\be
D_1 = v^2 \, m_{12}^2 \sin\beta \cos\beta\,,\quad
D_3 = {q^4 v^2 \over {4}}\cos^2\beta \left( m_{11}^2 - \sqrt{{\lambda_1 \over \lambda_2}} m_{22}^2\right)\left(\tan^2\beta - \sqrt{{\lambda_1 \over \lambda_2}}\right)\,,
\ee
Since we pay attention only to the signs of $D_i$, one can also redefine them by removing
factors which are guaranteed to be positive. With trivial manipulations one then arrives
at eq.~\eqref{eq:disc}.

\section{Classifying stationary points}
\label{ap:min}

Throughout this work we have used the method of Lagrange multipliers to determine
the existence of several stationary points. However, to discover whether they are
minima, maxima or saddle points, one must clearly look at the Hessian matrix computed
at each of the stationary points. In terms of the Lagrange multiplier $\zeta$, the Hessian
is easy to calculate, and is given by
\be
H \,=\, \left(
\begin{array}{cccc}
\Lambda_0 - \zeta & 0 & 0 & 0 \\
0 & \zeta - \Lambda_1 & 0 & 0 \\
0 & 0 & \zeta - \Lambda_2 & 0 \\
0 & 0 & 0 & \zeta - \Lambda_3  \\
\end{array}
\right)\;.
\label{eq:hes}
\ee
Recalling the results shown in section~\ref{sec:ast}, let us consider, for starters, the case
where $\Lambda_1 > \{\Lambda_2\,,\,\Lambda_3\}$. We have said that the global minimum occurs for
a value of the Lagrange multiplier $\Lambda_1 < \zeta_G < \Lambda_0$ - and indeed, we see from
the expression for the Hessian that with $\zeta = \zeta_G$ all the elements in the diagonal
of the matrix are positive, and as such this stationary point is guaranteed to be a minimum.
Likewise, the first stationary point - that with a value of $\zeta$ smaller than $\Lambda_3$ -
gives at least two negative values in the diagonal, and as such cannot be a minimum.

It is rather trickier to verify if one of the other stationary points is a
minimum, or a saddle point. In fact, it is not {\em necessary} that all of the diagonal
elements of the Hessian matrix of~\eqref{eq:hes} be positive for a stationary point to
be a minimum. One has to require a milder condition, namely that $H$ be positive
definite in the subspace tangent to the stationary point, in the space of the $r_\mu$.
A necessary and sufficient condition for the existence of a normal minimum becomes rather
elaborate, and we will not present it here, since it is not necessary for the purposes
of this paper. But we draw the attention of the reader to two very simple necessary conditions
that must be obeyed so that one has a normal minimum, in the softly broken $Z_2$ model. It is
simple to show that the squared masses of the charged and pseudoscalar
Higgses are given, in terms of the Lagrange multiplier $\zeta$ and in the basis where the
$\Lambda$ tensor is diagonal, by
\be
m^2_{H^\pm} \,=\, \zeta \,k^2\,v^2\;\;\; , \;\;\; m^2_A \,=\, (\zeta - \Lambda_2)\,k^2\,v^2
\,\;\; .
\ee
Thus, we see that any normal minimum must correspond to a value of a Lagrange multiplier
which satisfies $\zeta> 0$ and $\zeta > \Lambda_2$. This latter condition, by the by, is the
reason why we had to consider that $\Lambda_2$ was not the largest of the $\Lambda_i$. However,
to specify that a given stationary point is a minimum, we would also need to look at the
squared masses of the CP-even scalars and require that they be positive - and that condition
cannot be cast into a simple form in terms of the parameters of the potential.

\end{document}